\documentclass{article}

\usepackage{arxiv}

\usepackage[utf8]{inputenc} 
\usepackage[T1]{fontenc}    
\usepackage{url}            
\usepackage{booktabs}       
\usepackage{amsfonts}       
\usepackage{amsmath}
\usepackage{nicefrac}       
\usepackage{microtype}      
\usepackage{graphicx}
\usepackage{doi}
\usepackage{caption,subcaption}
\usepackage{xcolor}
\usepackage{xfrac}
\usepackage{afterpage}
\newcommand{\JY}[1]{{\color{black}{#1}}}
\newcommand{\MK}[1]{{\color{black}{#1}}}
\newcommand{\GS}[1]{{\color{black}{#1}}}

\title{Modeling of 2D self-drifting flame-balls in Hele-Shaw cells}

\author{ {Jorge Yanez}\thanks{Corresponding author} \\
	Institute for Thermal Energy Technology and Safety\\
        Karlsruhe Institute of Technology\\
        Hermann-von-Helmholtz-Platz, 1\\
        Karlsruhe, Germany.\\
	\texttt{jorge.yanez@kit.edu} \\
	\And
	Leonid Kagan\\
	School of Mathematical Sciences,\\
        Tel Aviv University\\
        Tel Aviv, Israel.\\
	\texttt{kaganleo@tauex.tau.ac.il} \\
	\AND
	Gregory Sivashinsky \\
        School of Mathematical Sciences,\\
        Tel Aviv University\\
        Tel Aviv, Israel.\\
	\texttt{grishas@tauex.tau.ac.il} \\
	\And
        Mike Kuznetsov\\
        Institute for Thermal Energy Technology and Safety\\
        Karlsruhe Institute of Technology\\
        Hermann-von-Helmholtz-Platz, 1\\
        Karlsruhe, Germany.\\
	\texttt{mike.kuznetsov@kit.edu}
}
\date{}

\renewcommand{\shorttitle}{Modeling of 2D self-drifting flame-balls in Hele-Shaw cells.}

\begin{document}
\maketitle

\begin{abstract}
The disintegration of near limit flames propagating through the gap of Hele-Shaw cells has recently become a subject of active research. In this paper, the flamelets resulting from the disintegration of the continuous 
front a reinterpreted in terms of the Zeldovich 
flame-balls stabilized by volumetric heat losses. 
A complicated free-boundary problem for 2D self-drifting near circular flamelets is reduced to a 1D model. 
The 1D formulation is then utilized to obtain the locus of the flamelet velocity, radius, heat losses and Lewis numbers at which the self-drifting flamelet exists.

\end{abstract}


\section{Introduction}

In their book The Mathematical Theory of Combustion, Zeldovich et al. \cite{zeldovich1985mathematical} analysed theoretically the possibility of diffusively burning 3D spherical flame-balls. The occurrence of such configurations in 2D was specifically discarded. It was concluded that spherical flame-balls surrounded by motionless fluid are unstable. The authors then conjectured that flame-balls might be stabilized by radiative or convective heat losses. In the latter case, a motion would develop that in turn would modify the topology of the flame-ball which would assume the shape of a stocking-cap with a seemingly extinguished rear side.

Flame-balls were originally predicted in Zeldovich’s book Theory of Combustion and Detonation, Acad. Sci. USSR, 1944, translated to English in \cite{zeldovich1992selected}. Some years later, flame-balls were observed for the first time in microgravity, by Ronney and his collaborators, both in drop-tower experiments \cite{ronney1990near} and in the spacecraft \cite{ronney1998experimental}. Flame-balls were thereupon studied by Buckmaster et al. in a now classical series of theoretical papers \cite{buckmaster1990structure, buckmaster1991structure, buckmaster1993analytical, buckmaster1991flame, buckmaster1993influence} successfully refining the premises employed. Originally, a stationary spherical flame subjected to internal volumetric heat losses was considered \cite{buckmaster1990structure}. The quenching limits where obtained as well as a set of radii for which the resulting spherical structure is stable and could thus be observed. The original formulation \cite{buckmaster1990structure} was extended in \cite{buckmaster1991structure} to incorporate heat losses in the unreacted mixture, refining the stability domain. The stability issue was later comprehensively explored numerically by Bockhorn et al. \cite{bockhorn1999adaptive, gerlinger2000numerical, gerlinger2003numerical}. The assumption of no-advection in \cite{buckmaster1990structure, buckmaster1991structure} was relaxed in \cite{buckmaster1991flame}, where the effect of the Peclet number was also assessed. The premise of volumetric heat losses was extended to incorporate also the effect of conductive heat losses to a neighbouring wall \cite{buckmaster1993influence}. The theory developed in \cite{buckmaster1990structure, buckmaster1991structure}  was applied to hydrogen-air mixtures \cite{buckmaster1993analytical} subjected to radiative heat losses. The equivalence ratios ensuring the existence of flame-balls were obtained. Finally, Grcar \cite{grcar2009new} showed that stable cell-like flamelets subjected to convective (rather than radiative) heat losses are also possible.

An unconventional disintegration of near-limit low Peclet number flames evolving in the narrow gap of Hele-Shaw cells has recently become the subject of energetic experimental and numerical research \cite{veiga2020unexpected, escanciano2021characterization, martinez2019role}. Physically the issue is clearly related to the problem of unconfined 3D flame-balls \cite{buckmaster1990structure, buckmaster1991structure, buckmaster1993analytical, buckmaster1991flame, buckmaster1993influence}.

\begin{figure}[ht]
	\centering
	\includegraphics[width=0.6\textwidth]{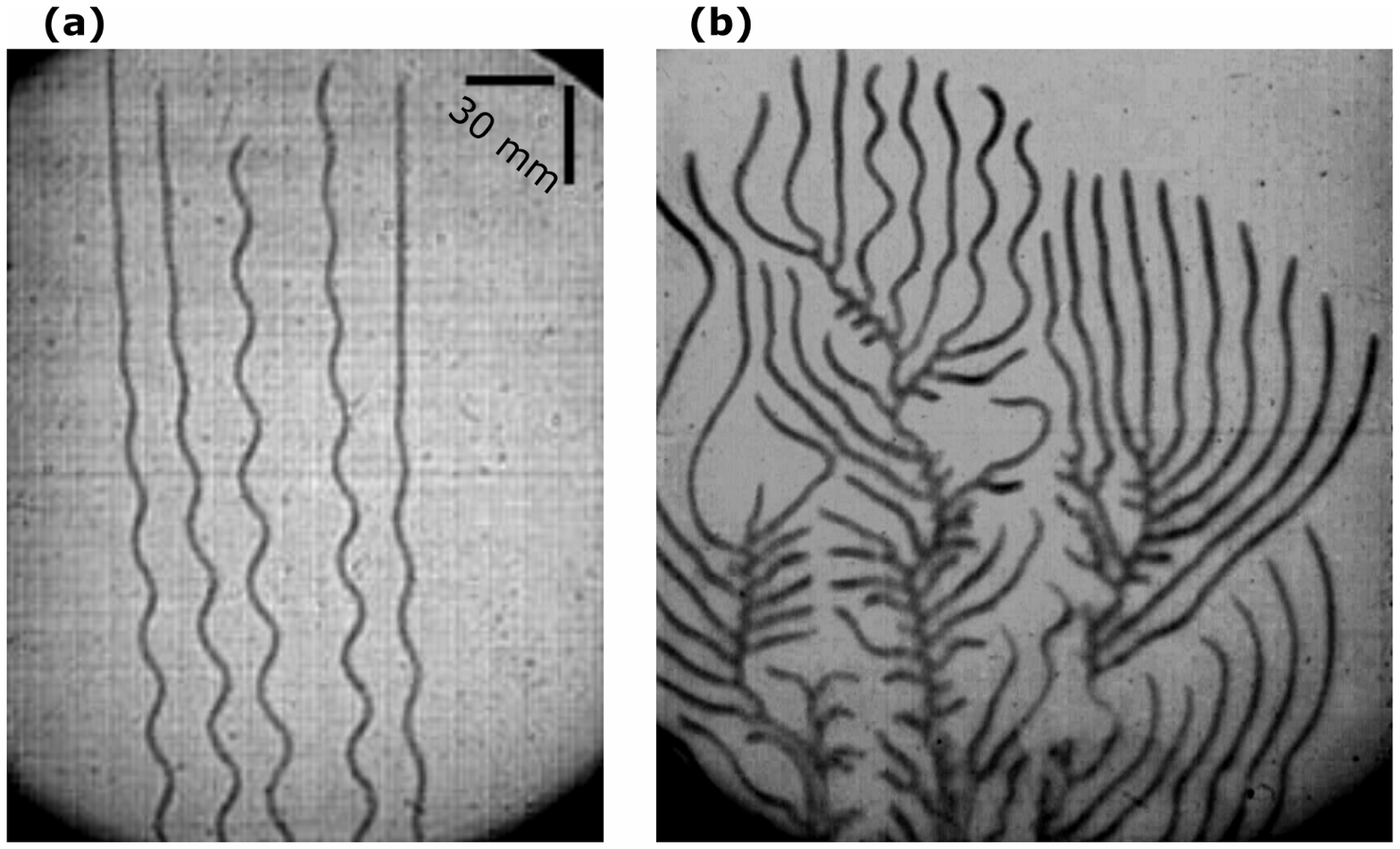}
	\caption{Monocellular hydrogen-air flamelets propagating through narrow gaps:
(a) one-headed finger, (b) one-headed branching.
Taken from Yanez et al. \cite{escanciano2021characterization}.}
	\label{fig:abb_gen2}
\end{figure}

Isolated flamelets propagating in a narrow gap between two parallel plates were photographed by Veiga et al. \cite{veiga2020unexpected} for Peclet numbers below 15. Among the regimes reported in \cite{veiga2020unexpected, escanciano2021characterization}  we focus in the present paper on the regimes shown in the Figure \ref{fig:abb_gen2}, in which flamelets propagate in a quasi-stationary regime.
 
\begin{figure}[ht]
	\centering\captionsetup{justification=centering}
	\includegraphics[width=0.3\textwidth]{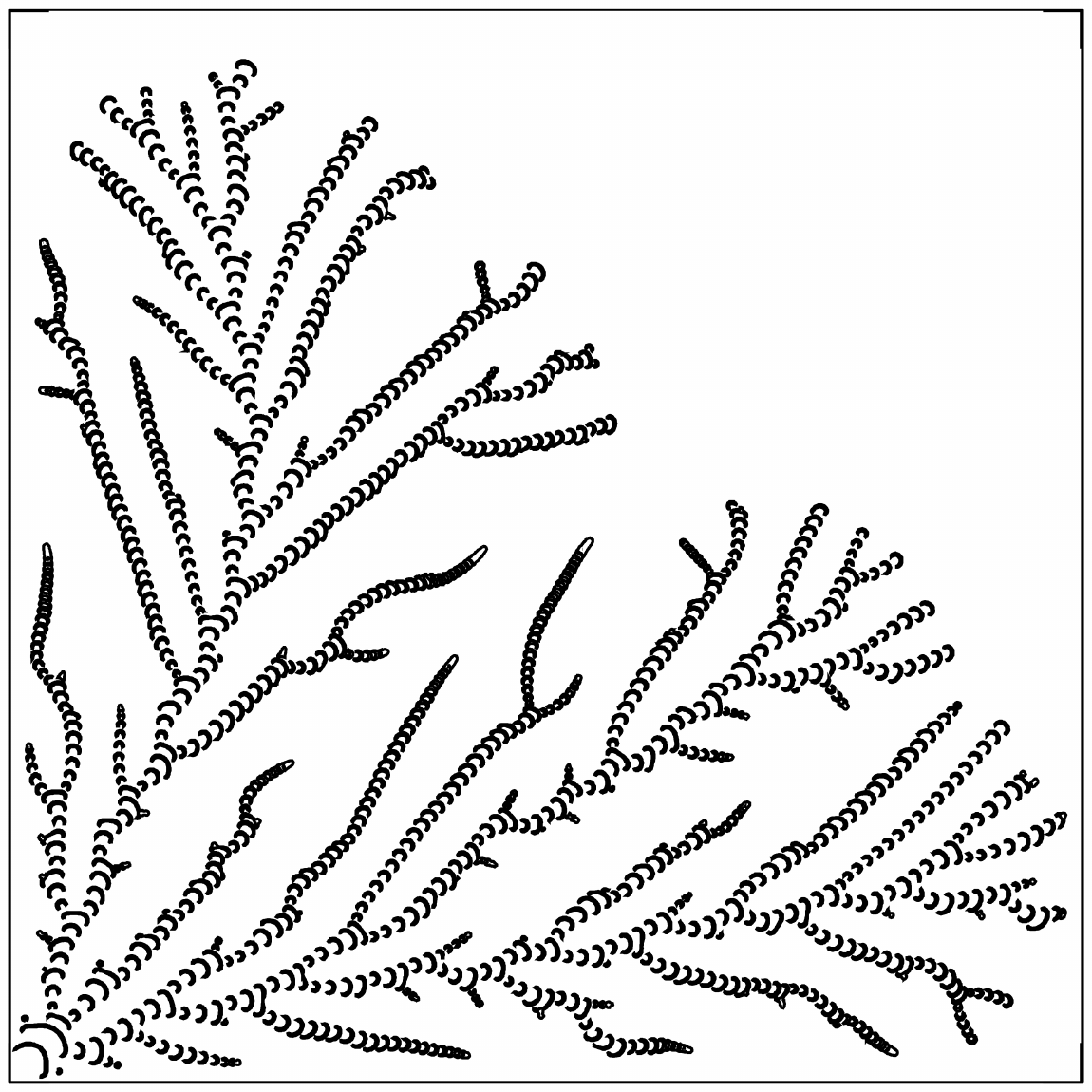}
\caption{ Successive monocellular flamelets under conductive heat losses. Simulation of the reaction-diffusion system [17] at $Q=q(T-\sigma)$ for $N=10,\;\sigma=0.2,\;Le=0.25,\;q=0.14$. The picture shows the reaction rate $\Omega$ distribution at several consecutive equidistant instants of time under initial conditions, $T(x,y,0)=(1-\sigma)\exp{[-(x^2+y^2)/l^2]}+\sigma$, $C(x,y,0)=1,\;l=0.25$, and adiabatic boundary conditions; $0<x,y<240$, $0<t<740$.}
	\label{fig:abb_gen3}
\end{figure}

\begin{figure}[ht]
	\centering
        \captionsetup{justification=centering}
	\includegraphics[width=0.3\textwidth]{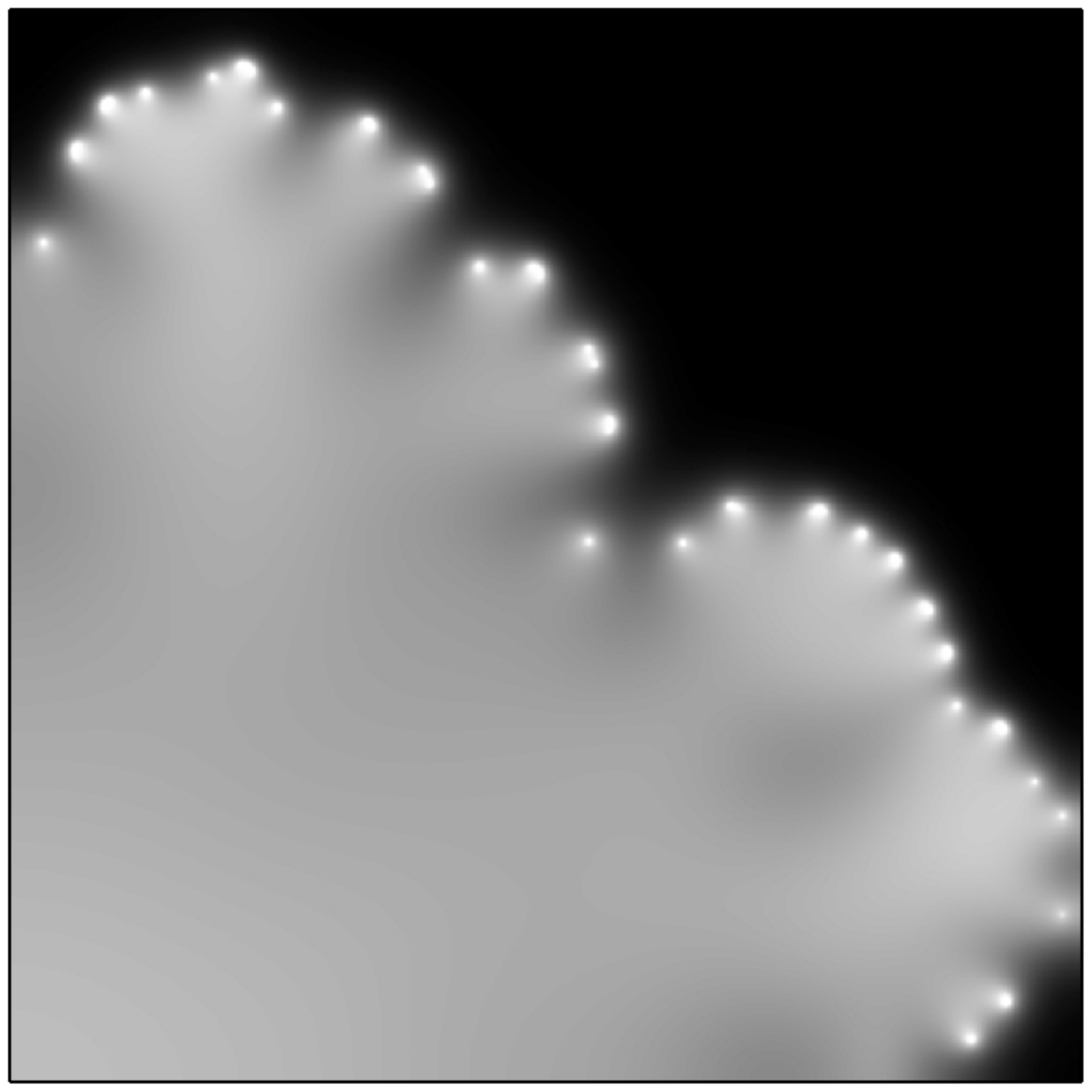}
	\caption{Distribution of the deficient reactant concentration	$C$ in 2D monocellular flamelets at $t=740$. The lighter regions correspond to lower concentration. For conditions see the legend for Fig. \ref{fig:abb_gen3}}
	\label{fig:abb_gen4}
\end{figure}

The shallow geometry of the Hele-Shaw cell suggests a 2D approach to the phenomenon. Kagan and Sivashinsky \cite{kagan1997self} studied the effect of volumetric heat losses on breaking up the flame in 2D. They assessed the changes of the front topology caused by different levels of heat losses. Some of the patterns observed are shown in Figures \ref{fig:abb_gen3}-\ref{fig:abb_gen4}. A 2D geometry was also utilized by Martinez-Ruiz et al. \cite{martinez2019role}  to reproduce numerically the flamelets, obtaining the stocking-cap flame shapes predicted by Zeldovich at al. \cite{zeldovich1985mathematical}.

There is an important distinction between the 3D and effectively the 2D situation typical of large aspect ratio Hele-Shaw cells. The 3D case allows for stationary spherical flame-balls which may bifurcate into a self-drifting mode \cite{brailovsky1997stationary, minaev2001self, minaev2002self, kagan2004self}. In the 2D case, the non-drifting circular flame-balls are ruled out since they cannot meet boundary conditions at infinity (due to the logarithmic tail of the associated concentration profiles \cite{martinez2019role}). So, in narrow gaps 2D self-drifting flame-balls should emerge not as a bifurcation but rather as the only way for the 2D flame-balls to exist.
Note that Ref. \cite{brailovsky1997stationary} was utilized in \cite{yanez2022velocity} to derive velocity, size and temperature of the flamelets appearing in the experiments of \cite{veiga2020unexpected} on Hele-Shaw cells.

Taking into account the considerations of the previous paragraphs, the objective of this paper is to ascertain the conditions that allow for self-drifting 2D flamelets subjected to conductive heat losses. Specifically, we will derive approximate expressions for the velocity, size and temperature of the flamelet. Conceptually, our study is a link between the flame topology suggested by Zeldovich et al. \cite{zeldovich1985mathematical} and the experimental findings of  \cite{veiga2020unexpected, escanciano2021characterization}. The present paper offers a simple 1D model for 2D self-drifting flame-balls.

\section{Model}
\subsection{General}

\begin{figure}[ht]
	\centering\captionsetup{justification=centering}
	\includegraphics[width=0.25\textwidth]{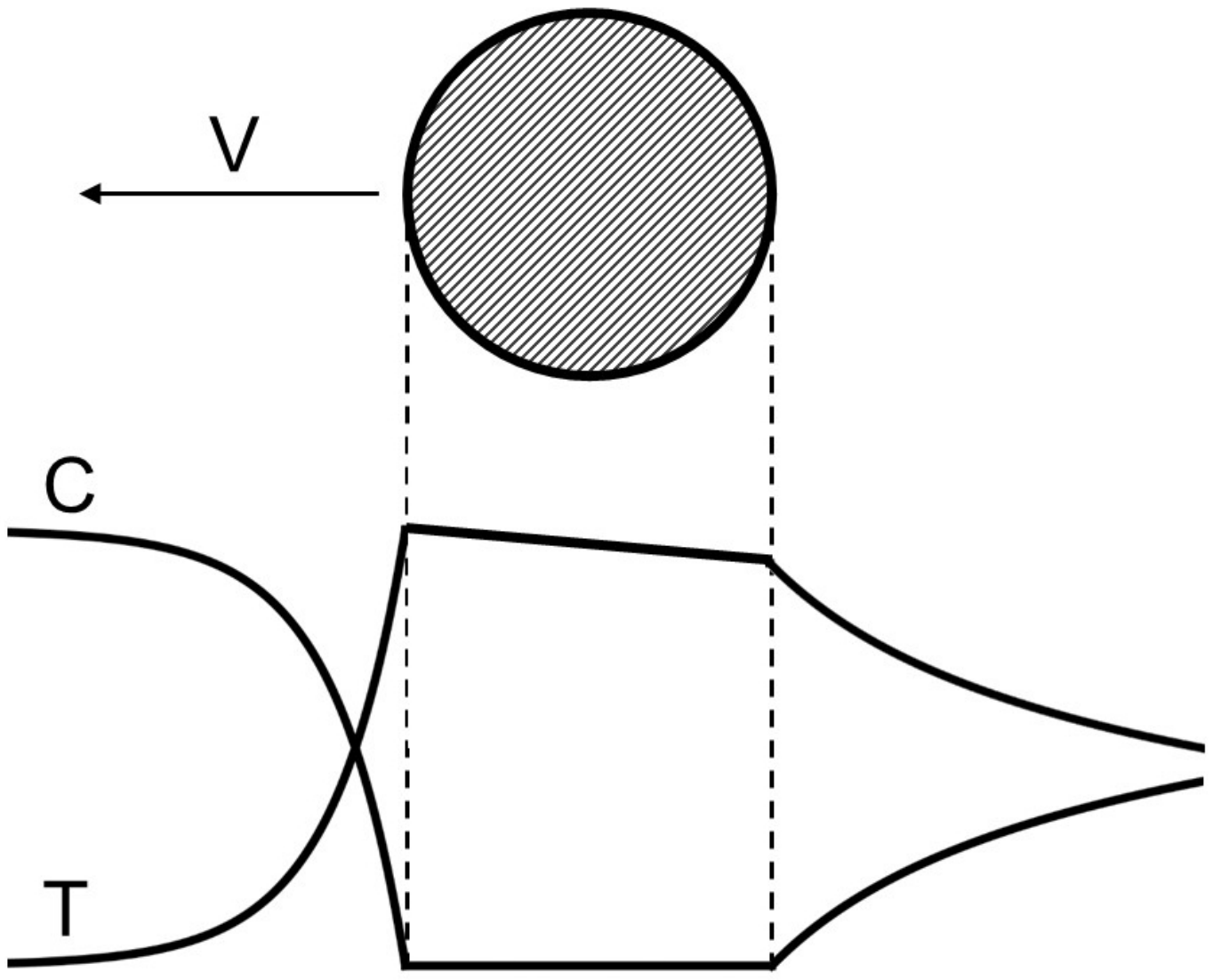}
	\caption{Up: Top sketch view of the flamelet in a plane parallel both to the paper and to the plates of the Hele-Shaw cell. Down: Profiles of temperature, $T$, and concentration, $C$, through the center of the flame-ball.}
	\label{fig:abb_gen}
\end{figure}

Let us consider the problem of a stationary near-circular flamelet of diameter $2R$ propagating in an infinite Hele-Shaw cell at a constant speed $V$. The reaction occurs within a thin interface preceded and followed by tails where diffusion of heat and reactants take place. The flamelet propagates from right to left (Figure \ref{fig:abb_gen}), so that in the reference frame attached to the flamelet the flow velocity is directed from left to right.

\subsection{Formulation}

\subsubsection{General considerations}

The mathematical model employed is based on the constant-density reaction-diffusion formulation \cite{kagan1997self, brailovsky1997stationary, yanez2022velocity, matkowsky1979asymptotic}. The flamelet velocity $V$ (Figure \ref{fig:abb_gen}) is regarded as an eigenvalue of the problem. The system of pertinent appropriately nondimensionalized equations read,
\begin{align}
	&\mathbf{V} \cdot \nabla T = \nabla^2 T + (1-\sigma)\Omega - Q,
        \label{eq:eq_T}\\
	&\mathbf{V} \cdot \nabla C = \frac{1}{Le}\nabla^2 C - \Omega, \label{eq:eq_C}
\end{align}

\JY{where $\mathbf{V}$ is the flow-field in units of $U_b$, the speed of a planar adiabatic flame; $T$ the temperature in units of $T_b$, the adiabatic temperature of combustion products; $C$ the concentration of the deficient reactant in units of $C_0$, its value in the fresh mixture; $\nabla=(\partial{}/\partial{x},\partial{}/\partial{y})$ the gradient
where $x,\;y$ are spatial coordinates in units of $l_{th}=D_{th}/U_b$, the thermal width of the flame; $D_{th} = $ thermal diffusivity of the mixture; $\sigma=T_0/T_b$, where $T_0$ is the temperature sustained at the walls of the Hele-Shaw cell. Additionally, $Q=q(T-\sigma)$ represents the scaled volumetric heat loss; $q$ stays for the scaled heat loss intensity specified as $(\pi l_{th}/d)^2$, where $d$ the width of the Hele-Shaw gap; $Le=D_{th}/D_{mol}$ is the Lewis number, where $D_{mol}$ is the molecular diffusivity of the deficient reactant; $N=T_a/T_b$ is the scaled activation energy, where $T_a$= activation temperature; $\Omega(C,T)=\mbox{\small $\frac{1}{2}$}(1-\sigma)^2Le^{-1}N^2C
 \exp{[N(1-T^{-1})]}$ is the appropriately normalized reaction rate to ensure
that at large $N$ the scaled speed of the planar adiabatic flame is close
to unity.} In polar coordinates the operators  $\nabla$, $\nabla^2$ read,
\begin{align}
	&\nabla \square=\cos{\theta}\dfrac{\partial \square}{\partial r} -\dfrac{\sin{\theta}}{r}\dfrac{\partial \square}{\partial \theta},\\
	&\nabla^2 \square = \dfrac{\partial^2\square}{\partial r^2} +\dfrac{1}{r}\dfrac{\partial\square}{\partial r} + \dfrac{1}{r^2}\dfrac{\partial^2 \square}{\partial\theta^2},
\end{align}
where $\square$ stays for $T$ or $C$. The thermo-diffusive model of \cite{matkowsky1979asymptotic} assumes a large activation energy, $N$. This simplification allows to treat the reaction zone as an infinitely thin interface, characterized by the locus $r=R(\theta)$. As a result, the reaction rate adopts a simple expression,
\begin{align}
	\Omega=&\sqrt{1+\dfrac{1}{R^2}\left(\dfrac{\partial R}{\partial \theta}\right)^2}
 \exp{\left(\dfrac{N}{2}\left(1-\dfrac{1}{T_f}\right)\right)} \delta\left(r-R\left(\theta\right)\right),
 \label{eq:rere}
\end{align}
 where $\delta$ is the Dirac delta function, and $T_f=T\left(r=R(\theta)\right)$ is the temperature at the flame-ball interface. 

\MK{In Eq. \eqref{eq:rere} $R^{-2}(\partial{R}/\partial{\theta})^2=(\partial{R}/\partial{s})^2$ where $ds=Rd\theta$ is the arc-length element of the flame interface, $r=R(\theta)$. Reduction of the reaction-diffusion system \eqref{eq:eq_T}-\eqref{eq:eq_C} to a free-boundary problem, valid for $N\gg 1$, is a widely used procedure in premixed flame theory (see e.g. \cite{matkowsky1979asymptotic}).}

\subsubsection{Boundary conditions}

 For the problem considered, the boundary conditions at $r\to \infty$ read,
 \begin{equation}
 	C\to 1, \;\,\; T\to\sigma.
 \end{equation}
For $r\le R(\theta)$,
 \begin{equation}
	C\equiv 0.
 \label{eq:Cinnerbc}
\end{equation}
At the flame interface, $r=R(\theta)$, $T=T_f(\theta)$, $C=0$,
\begin{align}
\left[\dfrac{\partial T}{\partial n}\right]_-^+&=\dfrac{1-\sigma}{Le}\left[\dfrac{\partial C}{\partial n }\right]_-^+,\\
\dfrac{1}{Le}\left[\dfrac{\partial C}{\partial n }\right]_-^+&=-\exp{\left(\dfrac{N}{2}\left(1-\dfrac{1}{T_f}\right)\right)}, 
\end{align}
where the normal derivative in polar coordinates reads,
\begin{equation}
	\dfrac{\partial \square}{\partial n} = \dfrac{\dfrac{\partial \square}{\partial r}-\dfrac{1}{R^2}\dfrac{\partial R }{\partial \theta}\dfrac{\partial \square}{\partial \theta}}{\sqrt{1+\dfrac{1}{R^2}\left(\dfrac{\partial R}{\partial \theta}\right)^2}}.
\end{equation}
Note that $T_f(\theta)$ is an unknown function to be determined.
\subsection{Solutions}
In the formulation adopted 
outside the flame front interface, $r = R(\theta)$, Eqs. \eqref{eq:eq_T}-\eqref{eq:eq_C} may be presented in the following unified form,
\begin{align}
&	
 \dfrac{\partial^2\square}{\partial r^2} +\dfrac{1}{r}\dfrac{\partial\square}{\partial r} + \dfrac{1}{r^2}\dfrac{\partial^2 \square}{\partial\theta^2}
	= 
 \nonumber\\&
	2 k \left[
	\cos{\theta}\dfrac{\partial \square}{\partial r} -\dfrac{\sin{\theta}}{r}\dfrac{\partial \square}{\partial \theta}
	\right]
	-
	q(T-\sigma).
	\label{eq:eq_gen}
\end{align}
Here $2k=V$ for $T$-equation and $q=0$, $2k=V Le$ for $C$-equation. 
Solutions of equation \eqref{eq:eq_gen} are known \cite{lamb1924hydrodynamics, tomotika1950steady}. 
Outside the flame-ball, at $q=0$, 
\begin{equation}
	\square = \sum_m A_m \exp{\left(k r \cos{\theta}\right)}\cos{m\theta}K_m\left(k r\right)+F\label{eq:expan1},
\end{equation}
while inside the flame-ball, at $q=0$,
\begin{equation}
	\square = \sum_m B_m \exp{\left(k r \cos{\theta}\right)}\cos{m\theta}I_m\left(k r\right) +G
 \label{eq:expan2},
\end{equation}
where $K_m$, $I_m$ are the modified Bessel functions of first and second kind and $A_m$, $B_m$, $F$, $G$ are parameters to be determined by the boundary conditions. Particular solutions pertinent to $q>0$ will be treated later.

\subsection{Simplification and rescaling}

For further analysis it is convenient to make the following change of variables
and parameters,
\begin{align}
	T&=\sigma +(1-\sigma)\Theta,
 \\
	Q&=(1-\sigma)q\Theta,
 \\
	\beta&=(1-\sigma)N.
 \end{align}
 At  $(1-\sigma)\ll 1$ and $\beta =N(1-\sigma)\gg 1$,
 \begin{align}
	\dfrac{1}{2}N(1-\dfrac{1}{T_f})
	&\simeq \frac{1}{2}\beta(\Theta_f-1),
\end{align}
where $\Theta$, $q$ are the re-scaled temperature and heat losses and $\beta$ is the Zeldovich number. So, equations \eqref{eq:eq_T}-\eqref{eq:eq_C} become,
\begin{align}
	&V  \left(\cos{\theta}\dfrac{\partial \Theta}{\partial r} -\dfrac{\sin{\theta}}{r}\dfrac{\partial \Theta}{\partial \theta}\right)  = \nabla^2 \Theta + \Omega-q\Theta,\label{eq:eq_T3}\\
	&V  \left(\cos{\theta}\dfrac{\partial C}{\partial r} -\dfrac{\sin{\theta}}{r}\dfrac{\partial C}{\partial \theta}\right) = \dfrac{1}{Le}\nabla^2 C -\Omega,\label{eq:eq_C3}
 \\
	&
 \Omega=\sqrt{1+\dfrac{1}{R^2}\left(\dfrac{\partial R}{\partial \theta}\right)^2} 
 \exp{\left\{\frac{1}{2}\beta(\Theta_f-1)\right\}} \delta\left(r-R\left(\theta\right)\right).
\end{align}

\subsection{Near-equidiffusive (NED) approximation}

We continue the analysis adopting the NED  approximation \cite{matkowsky1979asymptotic}. Let us combine \eqref{eq:eq_T3} and \eqref{eq:eq_C3} to obtain,
\begin{align}
	&V  \left(\cos{\theta}\dfrac{\partial \left(\Theta + C -1\right)}{\partial r} -\dfrac{\sin{\theta}}{r}\dfrac{\partial \left(\Theta +C-1\right)}{\partial \theta}\right)  = \nonumber\\
	&\nabla^2 \left[\Theta+\dfrac{1}{Le}C-1\right] -q\Theta.\label{eq:eq_s0}
\end{align}
In the NED limit, Lewis number is represented as $Le=1-\alpha/\beta$ where $\beta$ is regarded as a large parameter and $\alpha$ is a parameter of order one. In this approximation it is convenient to introduce the \emph{scaled excess enthalpy,} 
\begin{equation}
	S
	=(\Theta+C-1)\beta. 
 \label{eq:exc_enth}
\end{equation}
	At $\beta \gg 1$ we set,
\begin{align}
	\frac{1}{Le} & = 1+\frac{\alpha}{\beta},\label{eq:eq_app1}
 \\
	\Theta & = \Theta^{(0)}+\frac{1}{\beta}\Theta^{(1)}+\ldots,\label{eq:eq_app2}
 \\
	C& = C^{(0)}+\frac{1}{\beta}C^{(1)}+\ldots,\label{eq:eq_app3}
 \\
	q& = \frac{1}{\beta}\nu,\label{eq:eq_app4}
\end{align}
where superscripts in \eqref{eq:eq_app2}-\eqref{eq:eq_app3} mark the order of the $\nicefrac{1}{\beta}$ expansions; $\nu$ is the heat loss parameter of order one. Accounting for equations \eqref{eq:eq_s0}-\eqref{eq:eq_app4} one obtains,
\begin{align}
&	V \left(\cos{\theta}\dfrac{\partial S}{\partial r} -\dfrac{\sin{\theta}}{r}\dfrac{\partial S}{\partial \theta}\right)   = \nabla^2 \left[S+\alpha C^{(0)}\right] -\nu \Theta^{(0)},\label{eq:eq_isis1}\\
&
	V \left(\cos{\theta}\dfrac{\partial \Theta^{(0)}}{\partial r} -\dfrac{\sin{\theta}}{r}\dfrac{\partial \Theta^{(0)}}{\partial \theta}\right)  = \nabla^2 \Theta^{(0)} +\Omega,\label{eq:eq_isis2}\\
 &
	V  \left(\cos{\theta}\dfrac{\partial C^{(0)}}{\partial r} -\dfrac{\sin{\theta}}{r}\dfrac{\partial C ^{(0)}}{\partial \theta}\right)   = \nabla^2 C ^{(0)} -\Omega.\label{eq:eq_isis3}
\end{align}
Considering the boundary conditions, equations \eqref{eq:eq_isis2} and \eqref{eq:eq_isis3} readily yield, 
\begin{equation}\label{eq:szerodef}
	\Theta^{(0)}+C^{(0)}=1.
\end{equation} 
This last expression allows rewriting \eqref{eq:eq_isis1} in terms of $S$ and $\Theta^{(0)}$. The system of equations \eqref{eq:eq_isis1}-\eqref{eq:eq_isis3} thus becomes,
\begin{align}
&	
 V \left(\cos{\theta}\dfrac{\partial S}{\partial r} -\dfrac{\sin{\theta}}{r}\dfrac{\partial S}{\partial \theta}\right)   = \nabla^2 \left[S-\alpha \Theta^{(0)}\right] -\nu\Theta^{(0)},\label{eq:eq_qsis1}\\
 &
	V \left(\cos{\theta}\dfrac{\partial \Theta^{(0)}}{\partial r} -\dfrac{\sin{\theta}}{r}\dfrac{\partial \Theta^{(0)}}{\partial \theta}\right)  = \nabla^2 \Theta^{(0)} +\Omega.\label{eq:eq_qsis2}
\end{align}
At $r=R(\theta)$ we have,
\begin{align}
	&
 \left[S\right]_-^+=0, \;\;\; \Theta^{(0)}=1,
 \\
	&
 \left[\dfrac{\partial S}{\partial n}\right]_-^+=\alpha	\left[\dfrac{\partial \Theta^{(0)}}{\partial n}\right]_-^+,
 \\
	&
 \left[\dfrac{\partial \Theta^{(0)}}{\partial n}\right]_-^+
 =
 -\sqrt{1+\dfrac{1}{R^2}
 \left(\dfrac{\partial R}{\partial \theta}\right)^2}
 \exp{\left(\frac{1}{2}S\right)}.
 \label{eq:eq_qsis5}
\end{align}
The problem is thus reformulated in terms of $S$ and $\Theta^{(0)}$ and parameters $\alpha$, $\nu$. It does not involve the Zeldovich number $\beta$. Inside the flame-ball, $r<R(\theta)$, one has $\Theta^{(0)}\equiv 1$. 

\subsection{Reduction to 1D model}

As it stands the formulated 2D free-boundary problem \eqref{eq:eq_qsis1}-\eqref{eq:eq_qsis5} is too difficult for a straightforward analytical treatment. We therefore propose its low-mode collocation-like reduction to a 1D model that we believe will keep enough contact with the original 2D formulation. 

In the 1D approach one deals with a single spatial variable $z$ which runs from $z = -\infty$ to $z=\infty$ and in which the interior of the flame-ball covers the interval $-R<z<R$. In the expansions \eqref{eq:expan1}-\eqref{eq:expan2} we will keep only the leading terms corresponding to $m=0$. Moreover, we will set $\theta=0$ for $z>0$ and $\theta=\pi$ for $z<0$.

Conditions on the flame-ball interface $z=\pm R$ then become $\left[\Theta^{(0)}\right]_-^+=\left[\Theta^{(1)}\right]_-^+=\left[S\right]_-^+=0$-continuity and the slope jumps,
\begin{align}
&\left[\dfrac{d \Theta^{(0)}}{dz}\right]_-^+ + \exp{\left(\dfrac{1}{2}S\right)}=0,\label{eq:totsca1}\\
&\left[\dfrac{d S}{dz}\right]_-^+ = \alpha \left[\dfrac{d \Theta^{(0)}}{dz}\right]_-^+.\label{eq:totsca2}
\end{align}
Moreover, we have the conditions at infinity, $\Theta(z=\pm \infty)=0$, $S(z=\pm \infty)=0$, $C(z=\pm \infty)=1$ and full consumption of the deficient reactant in the interior of the flame-ball, $C(-R<z<R)=0$.

\JY{In the simplification employed one neglects deviation of the flame-ball shape from 
ideal circle.  This deviation however is deemed to be small.  In the construction of an approximate solution, rather than working with Eqs. \eqref{eq:eq_qsis1}-\eqref{eq:eq_qsis5}, we truncate the series involving an infinite number of terms, as in Eqs. \eqref{eq:expan1}-\eqref{eq:expan2}, to deal with sums involving only a finite number of terms.

For a given $V$	the problem is over-determined due to the boundary condition $\Theta^{(0)}=1$ or $C^{(0)}$ at the front (see Eq. \eqref{eq:szerodef}).
It is solvable only for the specific relation between $V,\;\alpha$ and $\nu$.}

\subsection{Inner heat loss model}

It is instructive first to consider the mathematically simplest situation accounting for the heat loss only within the flame-ball interior, setting $q=0$ for $|z|>R$. As a result one ends up with the following system. For $\Theta^{(0)}$, 
\begin{align}
&
\Theta^{(0)}(z<-R)
=
\exp{\left(\frac{1}{2}V(z+R)\right)}
\dfrac{K_0\left(-\frac{1}{2}Vz\right)}{K_0\left(\frac{1}{2}VR\right)},
\label{eq:thepro1}
\\
&
\Theta^{(0)}(z>R)
=
\exp{\left(\frac{1}{2}V(z-R)\right)}
\dfrac{K_0\left(\frac{1}{2}Vz\right)}{K_0\left(\frac{1}{2}VR\right)},
\label{eq:thepro2}
\\
&
\Theta^{(0)}(-R<z<R)=1.
\label{eq:thepro3}
\end{align}

For $\Theta^{(1)}$ a similar procedure can be undertaken, considering the external solution \eqref{eq:expan1} applied to the equation for $S$, \eqref{eq:eq_qsis1}. Considering the definition of the \emph{excess enthalpy}  \eqref{eq:exc_enth}, zero order expansion \eqref{eq:szerodef} and the boundary conditions \eqref{eq:Cinnerbc} one readily obtains,
\begin{align}
\Theta^{(1)}(z>R)&=S_+\Theta^{(0)}(z>R),
\label{eq:theta111}
\\
\Theta^{(1)}(z<-R)&=S_-\Theta^{(0)}(z<-R),
\label{eq:theta112}
\\
\Theta^{(1)}(-R<z<R)&=S(-R<z<R),
\end{align}
where $S_+=\Theta^{(1)}(R)$ and $S_-=\Theta^{(1)}(-R)$. 

For $S$, as follows from its definition \eqref{eq:exc_enth}, and the zero order approximation \eqref{eq:szerodef},
\begin{align}
&
S(z>R)=\Theta^{(1)}(z>R) + C^{(1)}(z>R),
\label{eq:S_eqs1}
\\
&
S(z<-R)=\Theta^{(1)}(z<-R) + C^{(1)}(z<-R).
\label{eq:S_eqs2}
\end{align}
Inside the flame-ball, a simplified expression for $S$ can be constructed from the solution \eqref{eq:expan2} augmented with the heat loss term, 
\begin{align}
&S\left(-R<z<R\right)=
A+B\exp{\left(\dfrac{1}{2}Vz\right)}I_0\left(\frac{1}{2}V|z|\right) - \dfrac{\nu}{V}z,\label{eq:prof_s_inside}
\end{align}

where the constants,
\begin{align}
    A=&\frac{1}{2}\left(S_++S_-\right)
    -\frac{1}{2}B I_0
    \left(\tfrac{1}{2}VR\right) 
    \left(
    \exp{\left(\tfrac{1}{2}VR\right)}
    +
    \exp{\left(-\tfrac{1}{2}VR\right)}
    \right),
    \\
     B=&\dfrac{2\nu R+V\left(S_+-S_-\right)}
     {V I_0\left(\frac{1}{2}VR\right) \left(\exp{\left(\frac{1}{2}VR\right)}-\exp{\left(-\frac{1}{2}VR\right)}\right)}.
\end{align}
arise from the continuity conditions at $\pm R$. 
Relations for $C^{(1)}$, are based on the equations,
\begin{align}
&
C(z>R)=1-\exp{\left[\tfrac{1}{2}V Le(z-R)\right]}\dfrac{K_0\left(\frac{1}{2}V Le z\right)}{K_0\left(\frac{1}{2}V Le R\right)},\label{eq:eqck1}
\\
&
C(z<-R)=1-\exp{\left[\tfrac{1}{2}V Le(z+R)\right]}\dfrac{K_0\left(-\frac{1}{2}V Le z\right)}{K_0\left(\frac{1}{2}V Le R\right)}.\label{eq:eqck2}
\end{align}
Here equations \eqref{eq:eqck1}-\eqref{eq:eqck2} are similar to \eqref{eq:thepro1}-\eqref{eq:thepro2} where $V$ is replaced by $V Le$ and boundary conditions $\Theta^{(0)}(\pm R)=1$ are replaced by $C(\pm R)=0$. Considering $1/\beta$ expansions and Eq. \eqref{eq:eq_app1}, one obtains,
\begin{align}
&
    C^{(1)}(z>R)
    =
    -
    \dfrac{1}{2}V\alpha
    \left[
    \dfrac{K_1\left(\frac{1}{2}V z \right)}{K_0\left(\frac{1 }{2}V z\right)}z
        - 
    \dfrac{K_1\left(\frac{1}{2}V R \right)}{K_0\left(\frac{1}{2}V Rz \right)}R
    -
    (z-R)
    \right]
    \Theta^{(0)}(z>R),
    \label{eq:C111}
    \\   
    &
    C^{(1)}(z<-R)
    =
    \frac{1}{2}V\alpha
    \left[
    \dfrac{K_1\left(-\frac{1}{2}V z \right)}{K_0\left(-\frac{1}{2}V z \right)}z
            + 
    \dfrac{K_1\left(\frac{1}{2}V R \right)}{K_0\left(\frac{1 }{2}V Rz\right)}R
    +
    (z+R)
    \right]
    \Theta^{(0)}(z<-R),
    \label{eq:C112}
    \\ 
    &
    C^{(1)}(-R<z<R)=
    0.
\end{align}
Substituting the above relations into \eqref{eq:totsca1}-\eqref{eq:totsca2} one ends up with a system determining $S_+$, $S_-$, $V$ and $R$.

\subsection{Global heat losses}

Let us consider the equation \eqref{eq:eq_T3} for $\Theta$ at $r>R(\theta)$,
\begin{align}
	&2k  \left(\cos{\theta}\dfrac{\partial \Theta}{\partial r} -\dfrac{\sin{\theta}}{r}\dfrac{\partial \Theta}{\partial \theta}\right)  = \nabla^2 \Theta -q\Theta.\label{eq:eq_TT3}
 \end{align}
 For $m=0$ the solutions of \eqref{eq:eq_TT3} may be sought (see \eqref{eq:expan1}) as,
\begin{equation}
    \Theta=\exp{\left(k r \cos{\theta}\right)}f(r).
    \label{eq:The1dcomp}
\end{equation}
Substituting this into \eqref{eq:eq_TT3}, one obtains,
\begin{equation}
    \Theta=H\exp{\left(k r\cos{\theta}\right)} K_0 \left(\sqrt{k^2+q} r\right),\label{eq:The1dcomp2}
\end{equation}
$H$ being a constant. This  allows us to draw an important conclusion: the only effect of the heat loss term $q\Theta$ outside the flame-ball is the shrinking of the $K_0$-function, compared to the $q=0$ case.

For the 1D modeling adopted in this study Eq. \eqref{eq:The1dcomp2} implies,
\begin{align}
    \Theta(z>R)&=\left(1+\frac{1}{\beta}S_+\right)
    \exp{\left[k(z-R)\right]}\dfrac{K_0\left(\sqrt{k^2+q}z\right)}{K_0\left(\sqrt{k^2+q}R\right)},\label{eq:theta_global1}\\
    \Theta(z<-R)&=\left(1+\frac{1}{\beta}S_-\right)
    \exp{\left[k(z+R)\right]}\dfrac{K_0\left(-\sqrt{k^2+q}z\right)}{K_0\left(\sqrt{k^2+q}R\right)}.\label{eq:theta_global2}
\end{align}

\section{Locus of $V$, $R$, $\nu$}

\subsection{System of equations}

Four equations \eqref{eq:totsca1}-\eqref{eq:totsca2}
determine four unknowns of the problem, $S_+$, $S_-$, $V$, $R$. But since $\sfrac{\partial \Theta^{(0)}(-R^+)}{\partial z}=\sfrac{\partial \Theta^{(0)}(R^-)}{\partial z}=0$, equations \eqref{eq:totsca1} readily imply that,
\begin{align}
	S_+&=2 \ln{\left[-\dfrac{\partial \Theta^{(0)}(R^+)}{\partial z}\right]},\\
	S_-&=2 \ln{\left[\dfrac{\partial \Theta^{(0)}(-R^-)}{\partial z}\right]}.
\end{align}
Therefore, the number of equations and unknowns are reduced to two. Inside the flamelet, equation \eqref{eq:prof_s_inside} provides $S$,
so that finally,
\begin{align}
&
    \dfrac{dS(R^+)}{dz} 
    -
    \left\{\frac{1}{2}BV\exp{\left(\tfrac{1}{2}VR\right)}\left[I_0\left(\tfrac{1}{2}VR\right)
    +I_1\left(\tfrac{1}{2}VR\right)\right] 
    -\dfrac{\nu}{V}\right\}
    =\alpha\dfrac{d\Theta^{(0)}(R^+)}{dz},
    \label{eq:sys_neq1}\\
    &
    \dfrac{dS(-R^-)}{dz} 
    -
    \left\{\frac{1}{2}BV\exp{\left(-\tfrac{1}{2}VR\right)}\left[I_0\left(\tfrac{1}{2}VR\right)
    -I_1\left(\tfrac{1}{2}VR\right)\right] \right.
    \left.-\dfrac{\nu}{V}\right\}
    =\alpha\dfrac{d\Theta^{(0)}(-R^-)}{dz}.
    \label{eq:sys_neq2}
\end{align}
 
 To close this problem, we need to present expressions for $\sfrac{d\Theta^{(0)}(-R^-)}{dz}$, $\sfrac{d\Theta^{(0)}(R^+)}{dz}$, $\sfrac{dS(-R^-)}{dz}$ and $\sfrac{dS(R^+)}{dz}$. 
  
The derivatives of $\Theta^{(0)}$ can be obtained directly from equations \eqref{eq:thepro1}-\eqref{eq:thepro2}. 
Expressions for derivatives of $S$ are provided by Eqs. \eqref{eq:S_eqs1}-\eqref{eq:S_eqs2} and \eqref{eq:C111}-\eqref{eq:C112}. Here the formula for $\Theta^{(1)}$ depends on whether one considers the inner or global heat losses. For inner heat losses, $\Theta^{(1)}$ is specified by Eqs. \eqref{eq:C111}-\eqref{eq:C112}, whilst for global heat losses Eqs. \eqref{eq:theta_global1} and \eqref{eq:theta_global2} should be employed. In the latter case,
\begin{align}
&
\dfrac{d S(R^+)}{dz} = S_+ \dfrac{d \Theta^{(0)}(R^+)}{dz} -\Sigma,
\\
&
\dfrac{d S(-R^-)}{dz} = S_- \dfrac{d \Theta^{(0)}(-R^-)}{dz} +\Sigma,
\end{align}
where
\begin{align}
\Sigma 
=
&
\left(
    \dfrac{\nu}{V}+\dfrac{V\alpha}{2}
\right)
\left\{
    \dfrac{VR}{4}
    \left[
        \dfrac{2 K_1^2\left(\frac{1}{2}VR\right)-K_0^2\left(\frac{1}{2}VR\right)-K_0\left(\frac{1}{2}VR\right)K_2\left(\frac{1}{2}VR\right)}
        {K_0^2\left(\frac{1}{2}VR\right)}
    \right]
    +
    \dfrac{K_1\left(\frac{1}{2}VR\right)}{K_0\left(\frac{1}{2}VR\right)}
\right\}
+
\dfrac{V\alpha}{2},
\label{eq:laren}
\end{align}
and
\begin{align}
&
\dfrac{d\Theta^{(0)}(R^+)}{dz} = \dfrac{V}{2}\left[1-\dfrac{K_1\left(\nicefrac{1}{2} VR\right)}{K_0\left(\nicefrac{1}{2} VR\right)}\right],
\\
&
\dfrac{d\Theta^{(0)}(-R^-)}{dz} = \dfrac{V}{2}\left[1+\dfrac{K_1\left(\nicefrac{1}{2} VR\right)}{K_0\left(\nicefrac{1}{2} VR\right)}\right].
\end{align}
\JY{For the inner heat loss model $\nu$ in Eq. \eqref{eq:laren} should be set at zero.}

\subsection{Numerical procedure}
\label{sec:numproc}

In spite of its apparent complexity, for a fixed $\alpha$, \eqref{eq:sys_neq1}-\eqref{eq:sys_neq2} is a system of two equations with two unknowns, $R$, $V$, and a parameter $\nu$. For brevity let us denote the system of equations by $\mathbf{f}$ and the variables by $\mathbf{y}$.

The locii of solutions of complex non-linear systems may contain bifurcation or turning points that make naive direct solutions by direct successive application of root-finding algorithms ineffective. Therefore, the so-called \emph{continuation principle} \cite{seydel2009practical} has been utilized to obtain the --parametric-- solution. 

The procedure regards the solution as a function of a certain arc-length, $s$, which is constructed with the unknowns $y$ and the parameter $\nu$. A single --initial-- solution of the equations is then obtained utilising any root-finding algorithm, in our case Newthon-Raphson method which is readily available in Sympy \cite{10.7717/peerj-cs.103}. 
Considering the Jacobian of the equations, a tangent is traced which allows finding a nearby solution by a predictor-corrector methodology. Since the system,
\begin{equation}
0=\mathbf{df}=\mathbf{f_y} \mathbf{dy} +\mathbf{f_\nu}d\nu,
\end{equation}
is undetermined, it must be complemented with a so-called \emph{parametrization}. Among the different available formulations \cite{seydel2009practical}, we choose it to be
\begin{equation}
\sum_i dy_i^2+d\nu^2-ds^2=0,
\end{equation}
where $ds$ is an imposed arc-length increment, small enough to obtain the necessary accuracy and fulfill stability requirements. 

The so called \emph{direct method} we are utilizing requires the calculation of the extended Jacobian $[\mathbf{f_y}, \mathbf{f_\nu}]$. Such derivation is feasible for our system, since it has only two equations, two unknowns and one parameter. Nevertheless, the algebra involved is very extensive requiring the utilization of a symbolic algebra package, in our case, Sympy \cite{10.7717/peerj-cs.103}.

We refer the reader to the monograph of Seydel\cite{seydel2009practical} for further technical details.

\subsection{Results. Locus $V$-$R$-$\nu$-$\alpha$}

\begin{figure}[ht]
	\centering\captionsetup{justification=centering}
	\includegraphics[width=0.49\columnwidth]{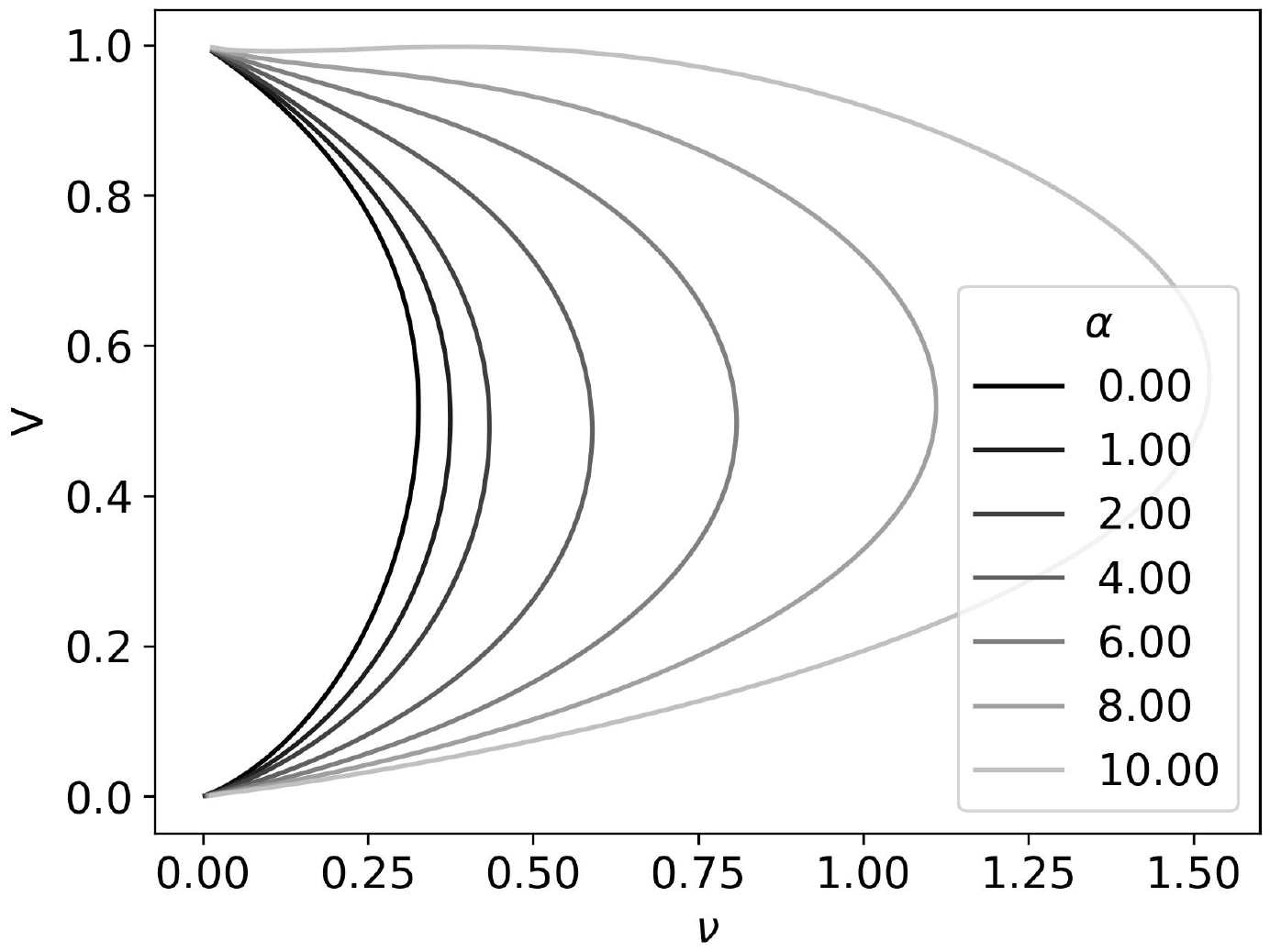}
	\caption{Locus of the velocities, $V$, as a function of the heat losses, $\nu$, for different Lewis number parameters, $\alpha$. Inner heat losses.}
	\label{fig:nuv1}
\end{figure}

\begin{figure}[ht]
	\centering\captionsetup{justification=centering}
	\includegraphics[width=0.49\columnwidth]{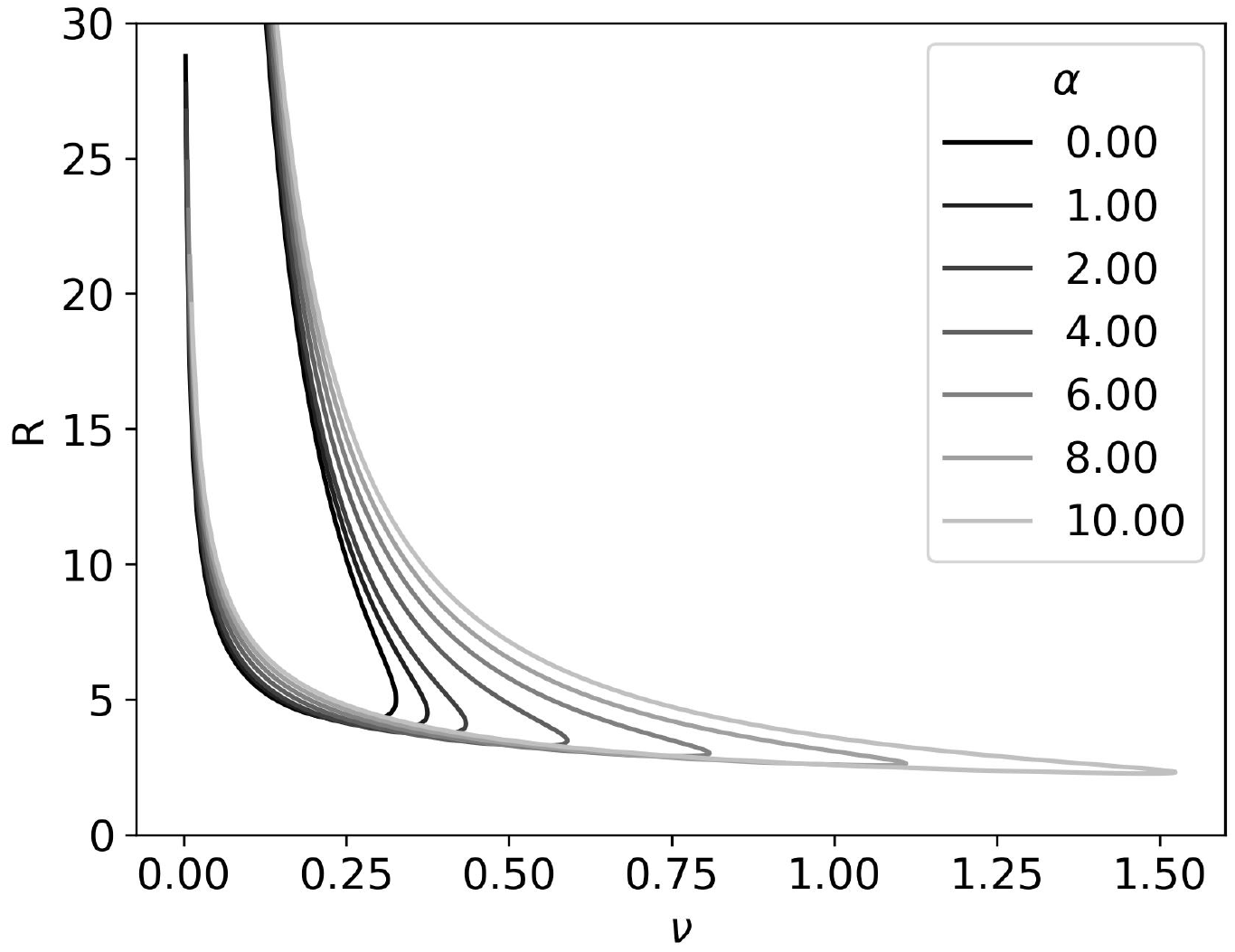}
	\caption{Locus of the flamelet radii, $R$, as a function of the heat losses, $\nu$, for different Lewis number parameters, $\alpha$. Inner heat losses.}
	\label{fig:nur1}
\end{figure}

\begin{figure}[ht]
	\centering\captionsetup{justification=centering}
	\includegraphics[width=0.49\columnwidth]{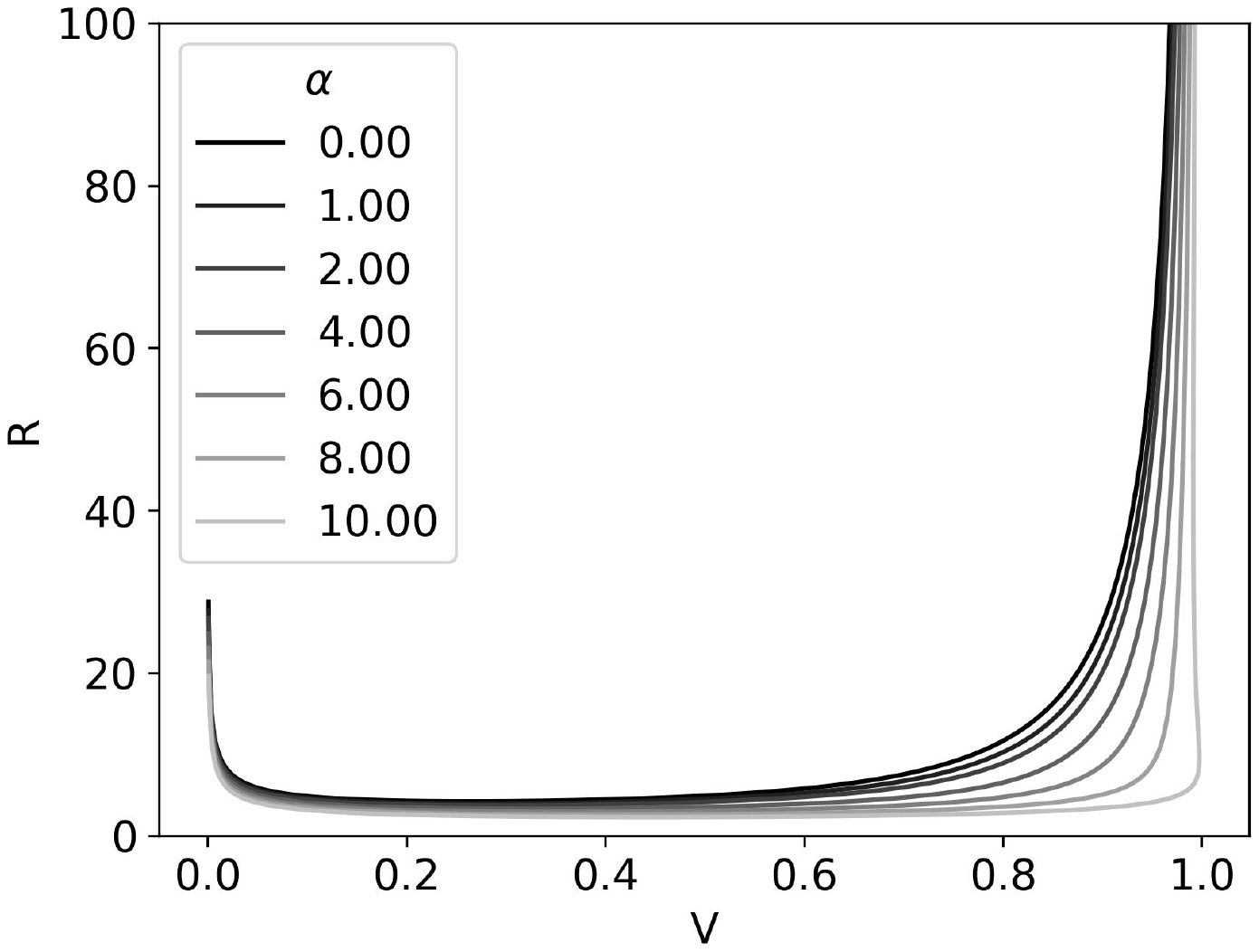}
	\caption{Locus of the flamelet radii, $R$, as a function of the velocities, $V$, for different Lewis number parameters, $\alpha$. Inner heat losses.}
	\label{fig:vr1}
\end{figure}

\begin{figure}[ht]
	\centering\captionsetup{justification=centering}
	\includegraphics[width=0.49\columnwidth]{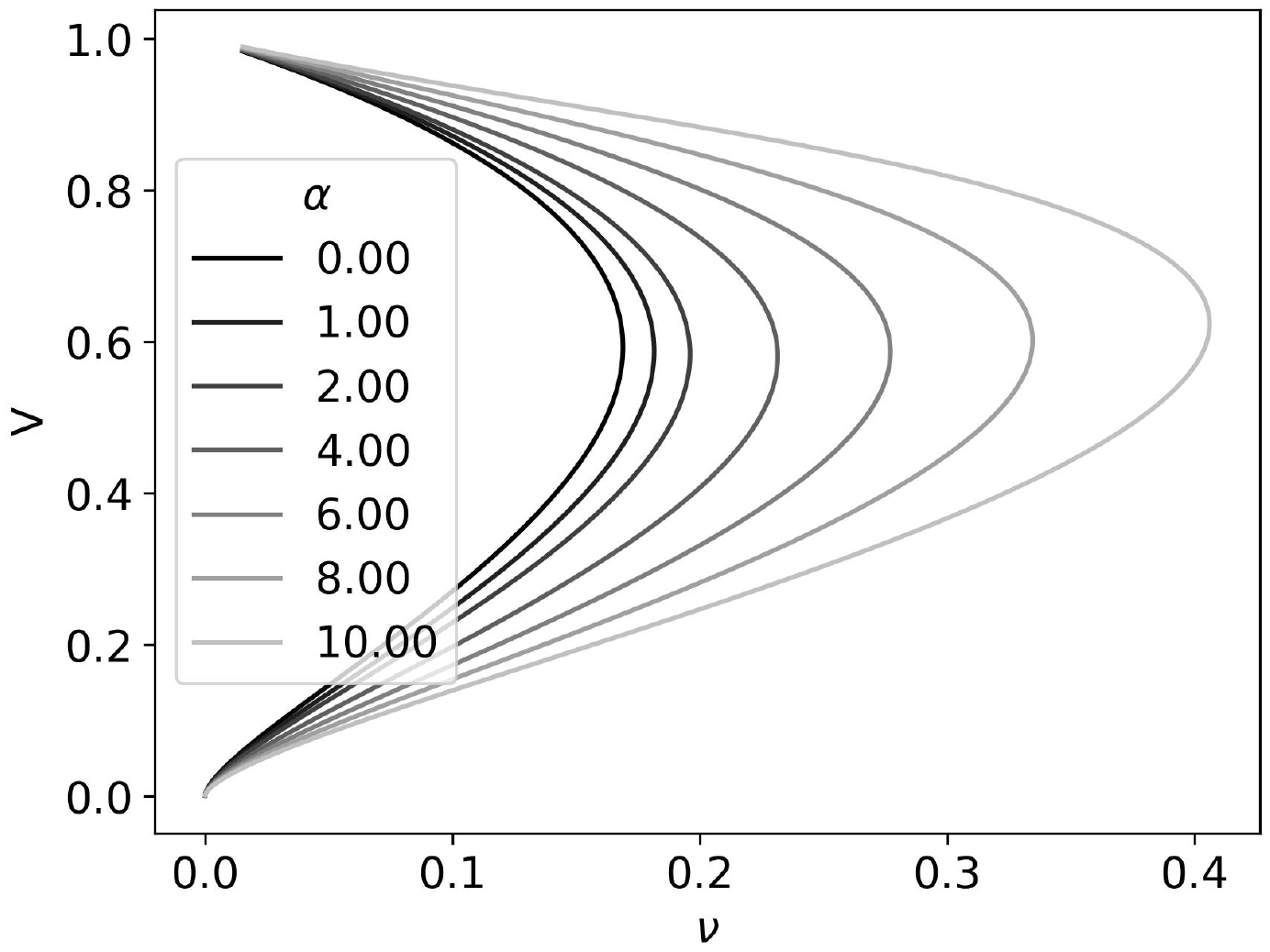}
	\caption{Locus of the velocities, $V$, as a function of the heat losses, $\nu$, for different Lewis number parameters, $\alpha$. Global heat losses.}
	\label{fig:nuv2}
\end{figure}

\begin{figure}[ht]
	\centering\captionsetup{justification=centering}
	\includegraphics[width=0.49\columnwidth]{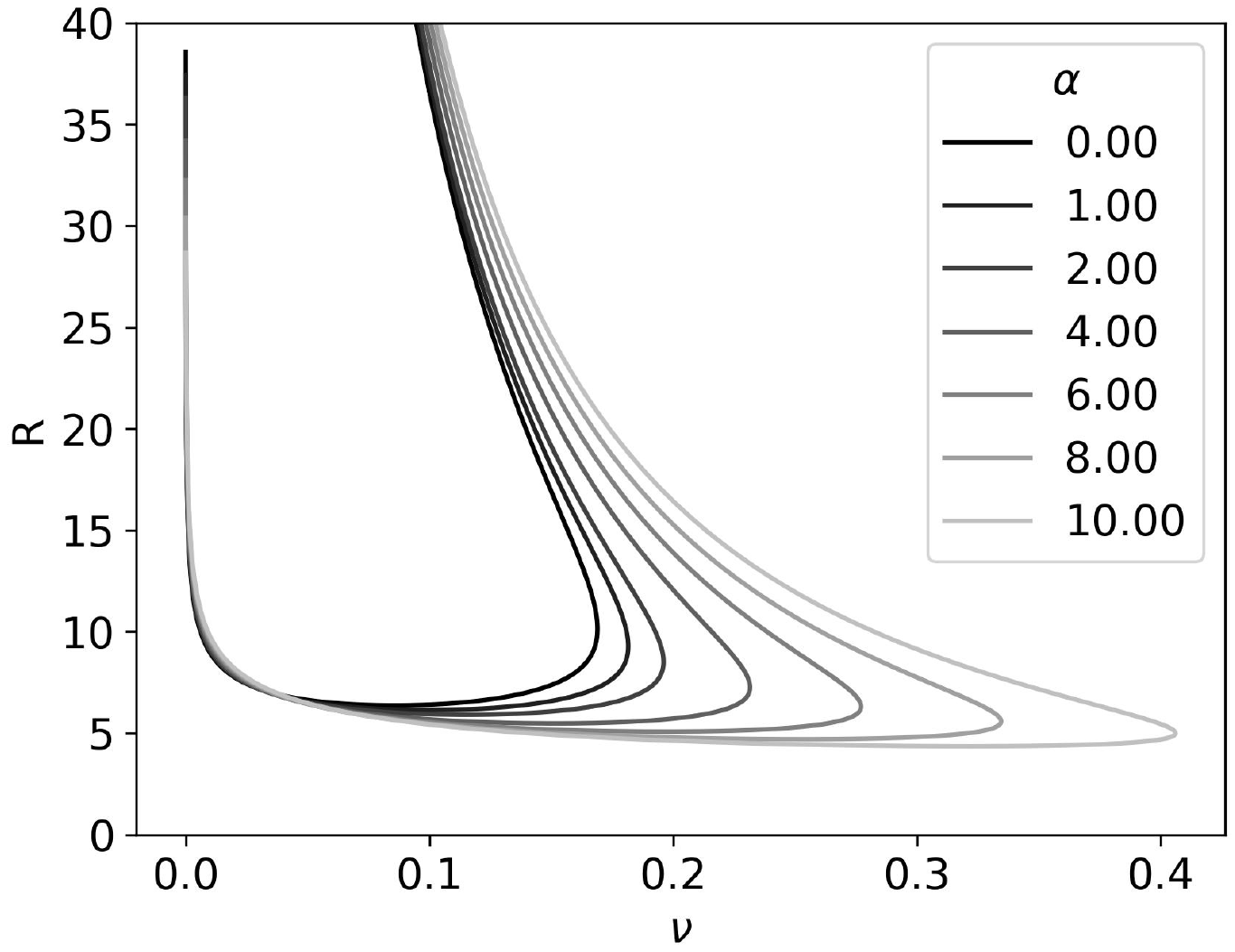}
	\caption{Locus of the flamelet radii, $R$, as a function of the heat losses, $\nu$, for different Lewis number parameters, $\alpha$. Global heat losses.}
	\label{fig:nur2}
\end{figure}

\begin{figure}[ht]
	\centering\captionsetup{justification=centering}
	\includegraphics[width=0.49\columnwidth]{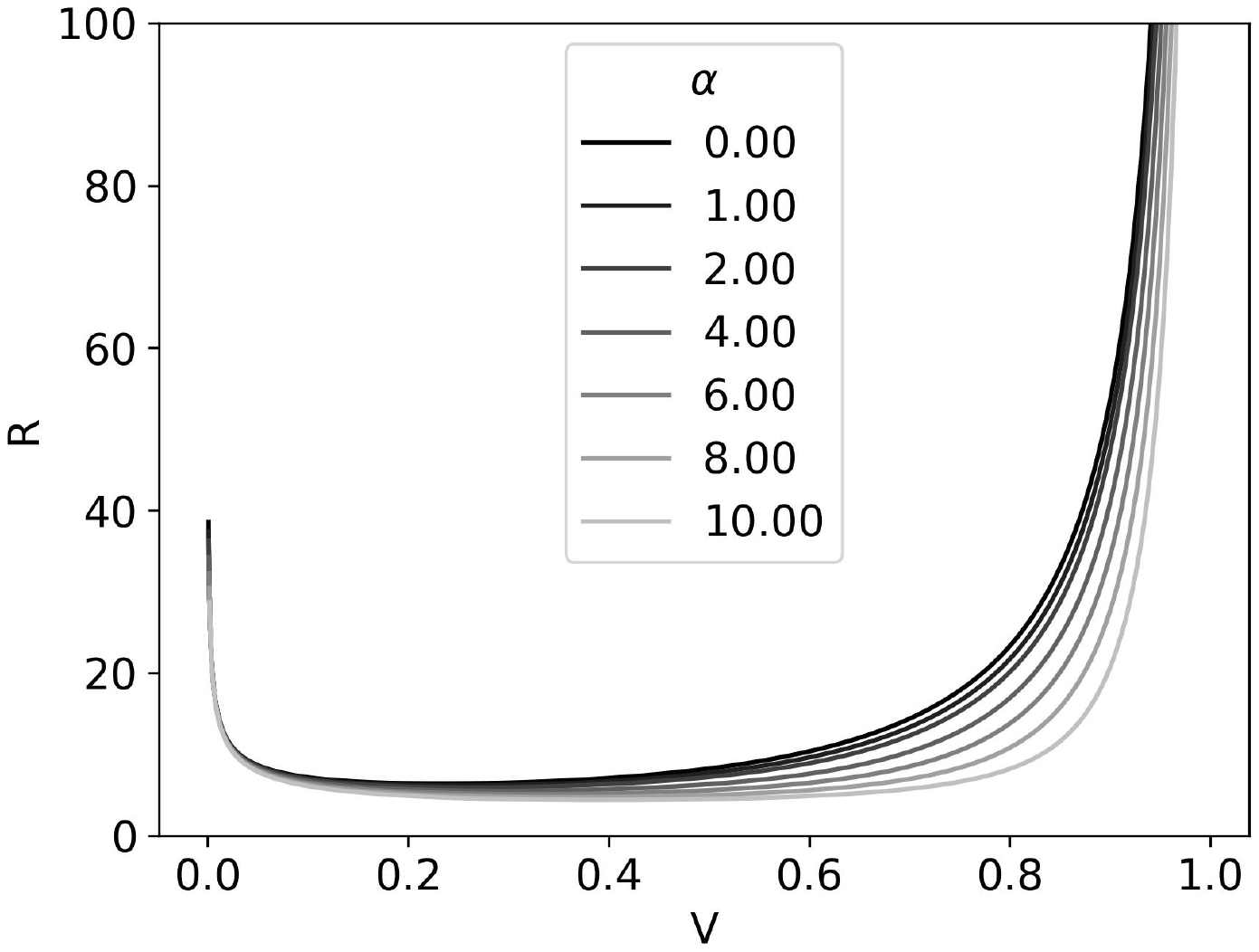}
	\caption{Locus of the flamelet radii, $R$, as a function of the velocities, $V$, for different Lewis number parameters, $\alpha$. Global heat losses.}
	\label{fig:vr2}
\end{figure}

The numerical procedure outlined in section \ref{sec:numproc} applied to equations \eqref{eq:sys_neq1}-\eqref{eq:sys_neq2} readily yields the dependencies shown in Figures \ref{fig:nuv1}, \ref{fig:nur1}, \ref{fig:vr1}, when the heat losses outside the flame-ball are suppressed. For global heat losses the pertinent diagrams are displayed in Figures \ref{fig:nuv2}, \ref{fig:nur2}, \ref{fig:vr2}.


One can draw several conclusions from the the results obtained. 
\begin{enumerate}
    \item Parameter $\alpha$ strongly affects $V(\nu)$, $R(\nu)$ dependencies. The heat loss intensity at the quenching (turning) point ($\nu_q$) increases with the decrease of the Lewis number, $Le=1-\nicefrac{\alpha}{\beta}$.

    \item Accounting for the heat loss outside the flame-ball ($|z|>R$) leads to a reduction of  $\nu_q$.
    
    \item 
    
\JY{At  $\nu<\nu_q$  the multiplicity of $V(\nu),\; R(\nu)$ dependencies is observed where only the regime with decreasing $V(\nu)$ is likely to be dynamically feasible. This would be in line both with numerical simulations \cite{martinez2019role} as well as recent experimental studies \cite{kuz2022kk}.

A similar non-uniqueness in known to occur in planar ($R=\infty$) non-adiabatic flames allowing for a comprehensive theoretical analysis which leads to the following $\alpha$-independent relation between $V$ and $\nu$ \cite{sivashinsky1981stability}, 
\begin{equation}
V^2\ln{\left(\frac{1}{V}\right)}=\nu.    
\end{equation}}

        \item A significant drop of temperature at the rear side $(z=R)$ of the self-drifting flame-ball might explain the stoking-cap shape of the advancing flamelet. The later is often perceived as a local extinction (opening) of the front.

\end{enumerate}


\subsection{Profiles across the flame-ball}

Once the solutions for $V(\nu)$, $R(\nu)$, for a certain $\alpha$ are available, we can plot  $\Theta^{(0)}$, $\Theta$ and $S$ vs. $z$. 

\begin{figure}[ht]
	\centering\captionsetup{justification=centering}
	\includegraphics[width=0.49\columnwidth]{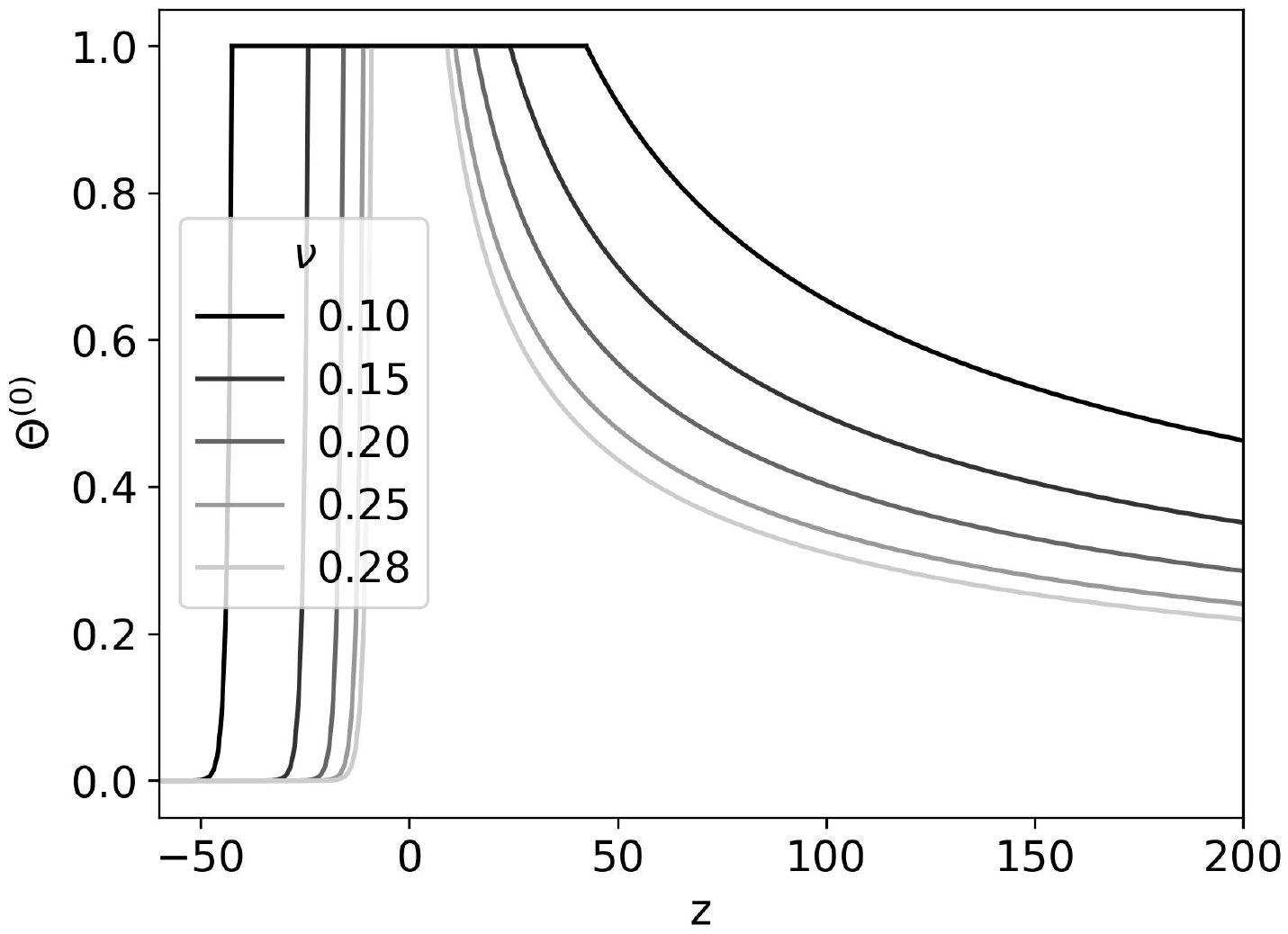}
	\caption{Profiles of $\Theta^{(0)}$ for $\alpha=1$ and different values of $\nu$ for inner losses. Upper branch.}
	\label{fig:teh1}
\end{figure}

\begin{figure}[ht]
	\centering\captionsetup{justification=centering}
	\includegraphics[width=0.49\columnwidth]{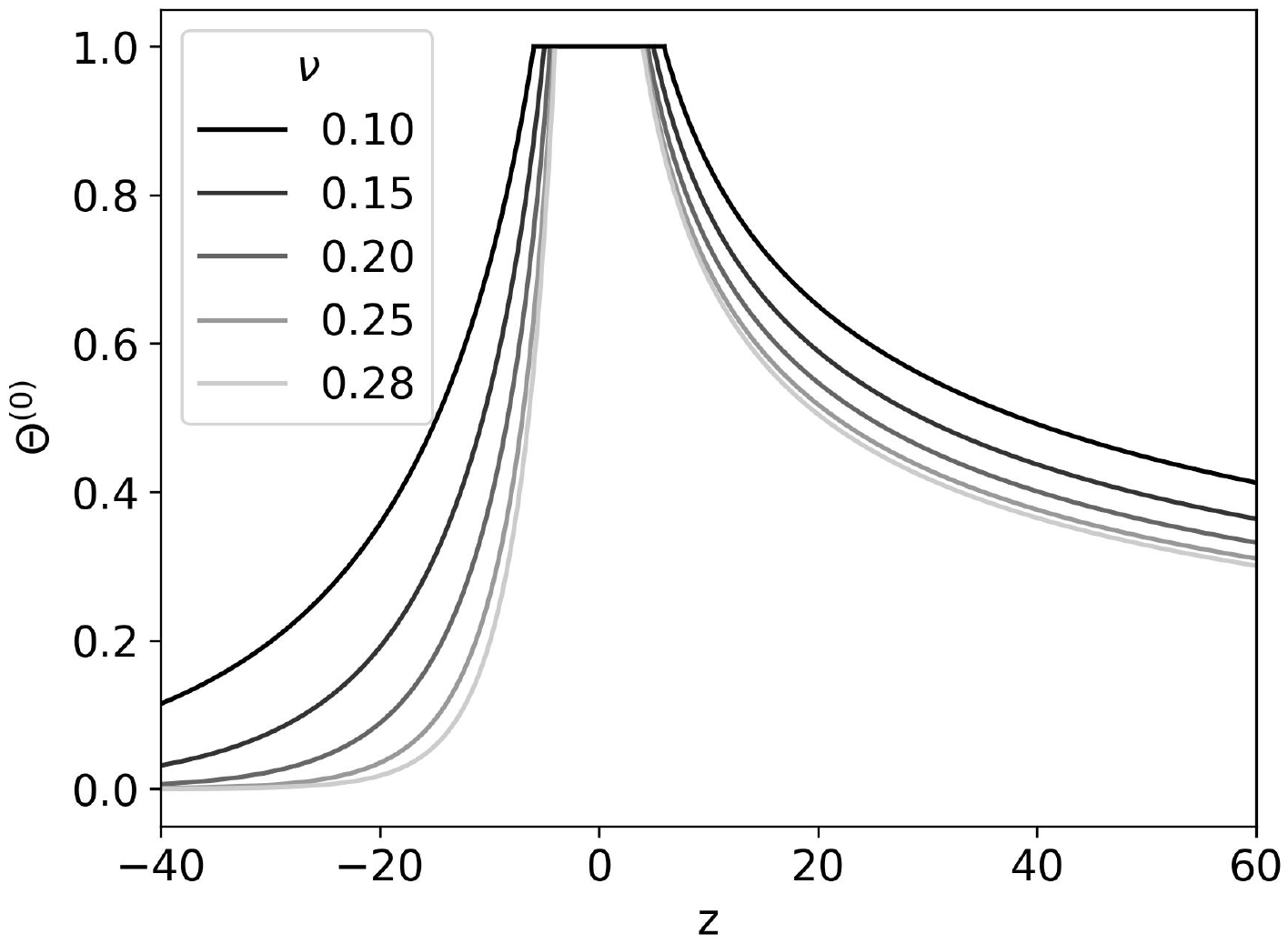}
	\caption{Profiles of $\Theta^{(0)}$ for $\alpha=1$ and different values of $\nu$ for inner losses. Lower branch.}
	\label{fig:teh1l}
\end{figure}

\begin{figure}[ht]
	\centering\captionsetup{justification=centering}
	\includegraphics[width=0.49\columnwidth]{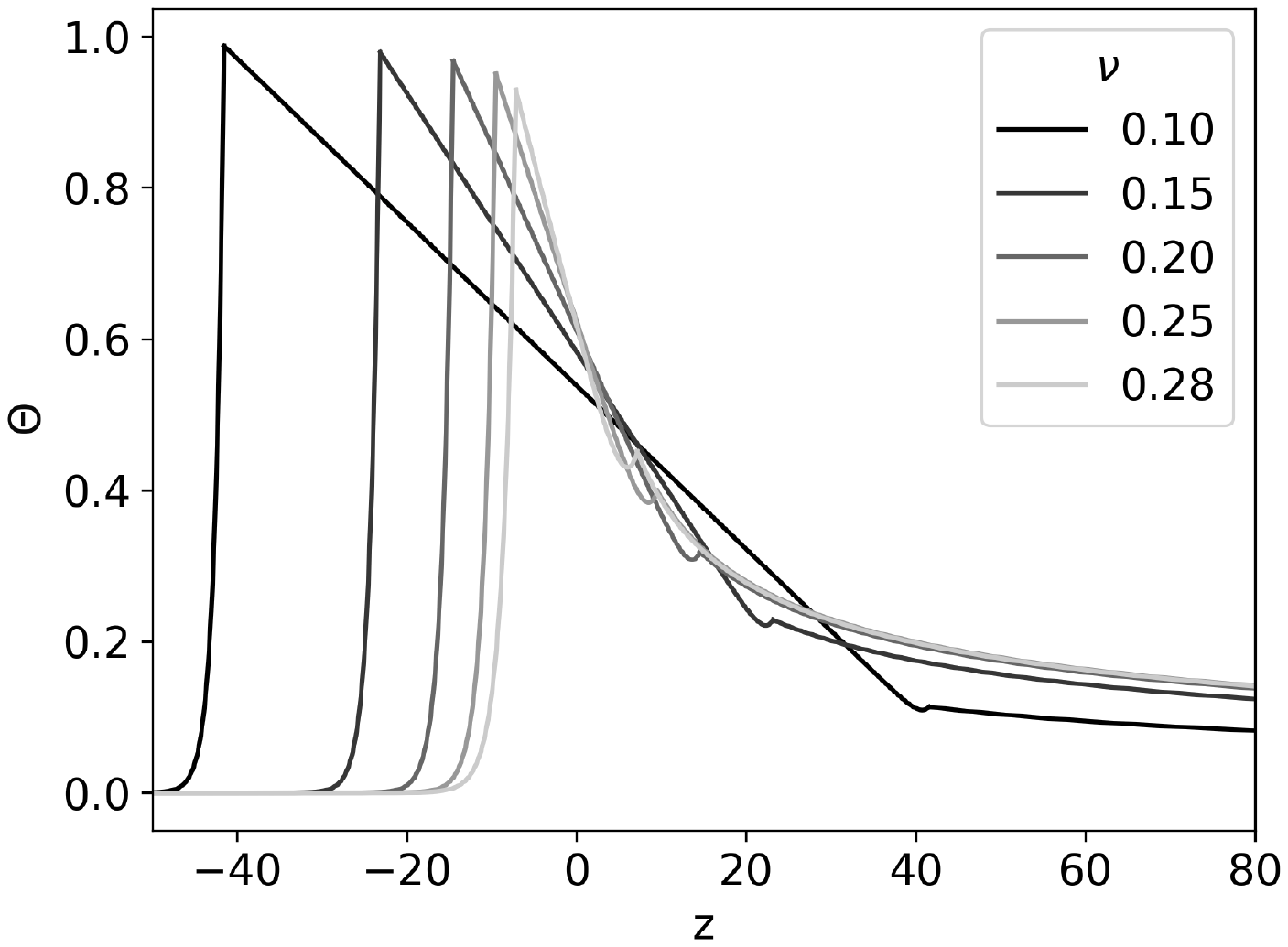}
	\caption{Profiles of $\Theta$ for $\alpha=1$, $\beta=10$ and different values of $\nu$ for inner losses. Upper branch.}
	\label{fig:teh1a}
\end{figure}

\begin{figure}[p]
	\centering\captionsetup{justification=centering}
	\includegraphics[width=0.49\columnwidth]{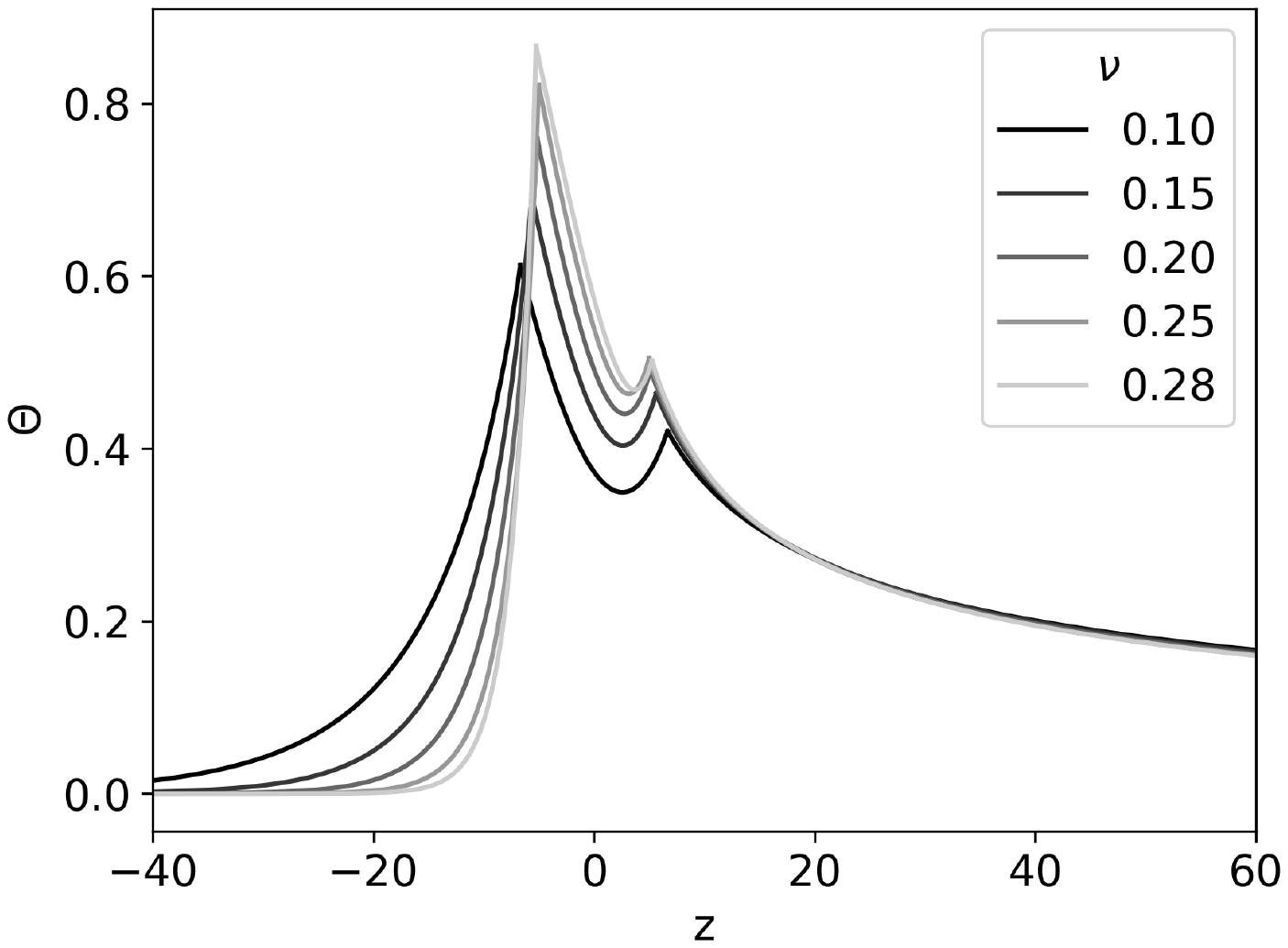}
	\caption{Profiles of $\Theta$ for $\alpha=1$, $\beta=10$ and different values of $\nu$  for inner losses. Lower branch.}
	\label{fig:teh1la}
\end{figure}

\begin{figure}[p]
	\centering\captionsetup{justification=centering}
	\includegraphics[width=0.49\columnwidth]{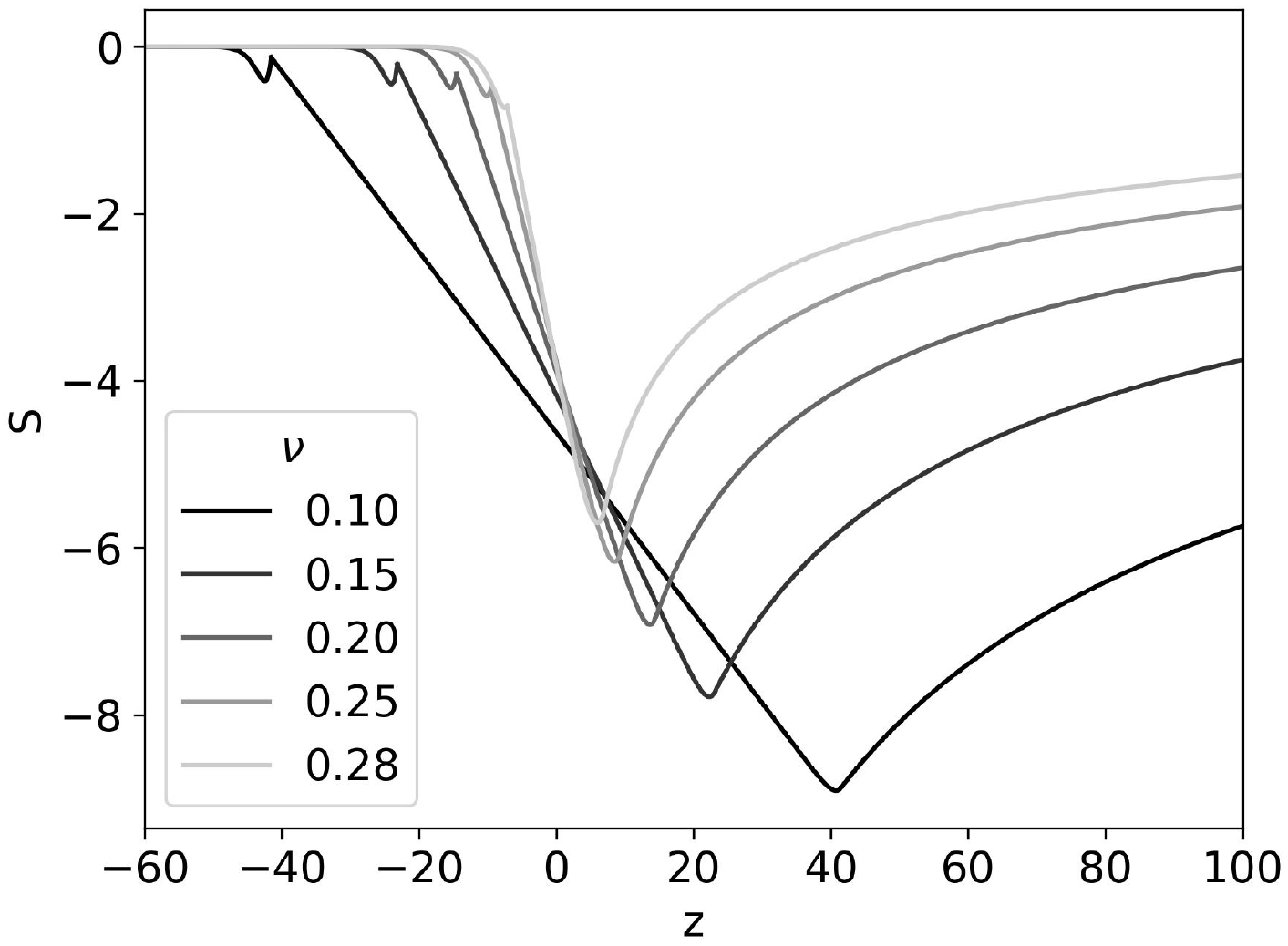}
	\caption{Profiles of $S$ for $\alpha=1$ and different values of $\nu$ for inner losses. Upper branch.}
	\label{fig:Sa}
\end{figure}

\begin{figure}[p]
	\centering\captionsetup{justification=centering}
	\includegraphics[width=0.49\columnwidth]{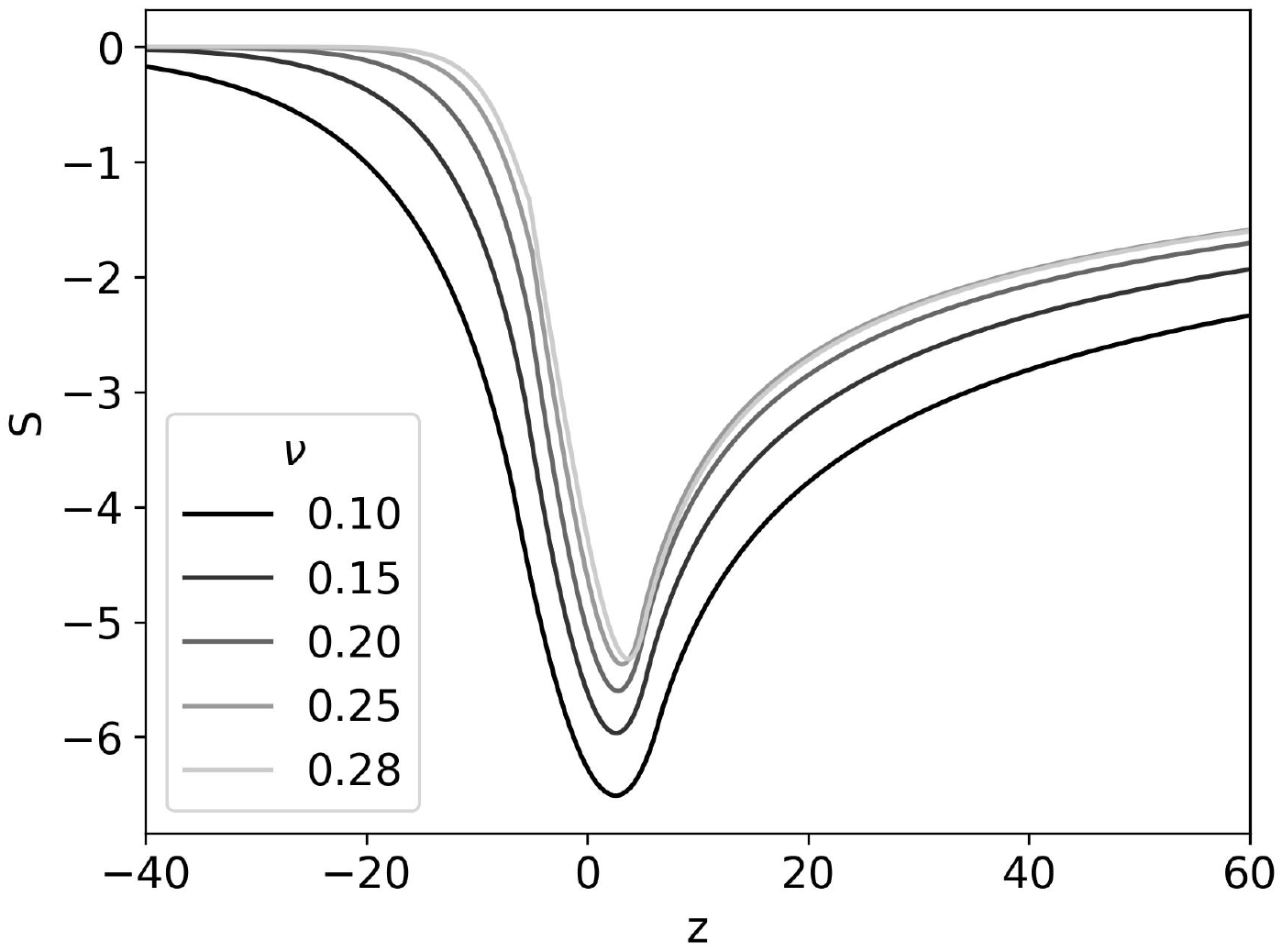}
	\caption{Profiles of $S$ for $\alpha=1$ and different values of $\nu$   for inner losses. Lower branch.}
	\label{fig:S1l}
\end{figure}

For the case in which the heat losses outside the flamelets are disregarded, Eqs. \eqref{eq:thepro1}-\eqref{eq:prof_s_inside} provide the required profiles. Figures \ref{fig:teh1} to \ref{fig:S1l} depict results for various heat losses at $\alpha = 1$. Figures \ref{fig:teh1}, \ref{fig:teh1a} and \ref{fig:Sa} show the profiles for the upper branch, whilst Figures \ref{fig:teh1l}, \ref{fig:teh1la} and \ref{fig:S1l} the profiles for the lower branch. 

In Figures \ref{fig:teh1}-\ref{fig:teh1l} one observes a steep slope of the curves on the left compared to a relatively slow decay on the right. 

A much more realistic temperature profile arises considering higher orders of the $1/\beta$-expansions, Figures  \ref{fig:teh1a}-\ref{fig:teh1la}. 

\begin{figure}[p]
	\centering\captionsetup{justification=centering}
	\includegraphics[width=0.49\columnwidth]{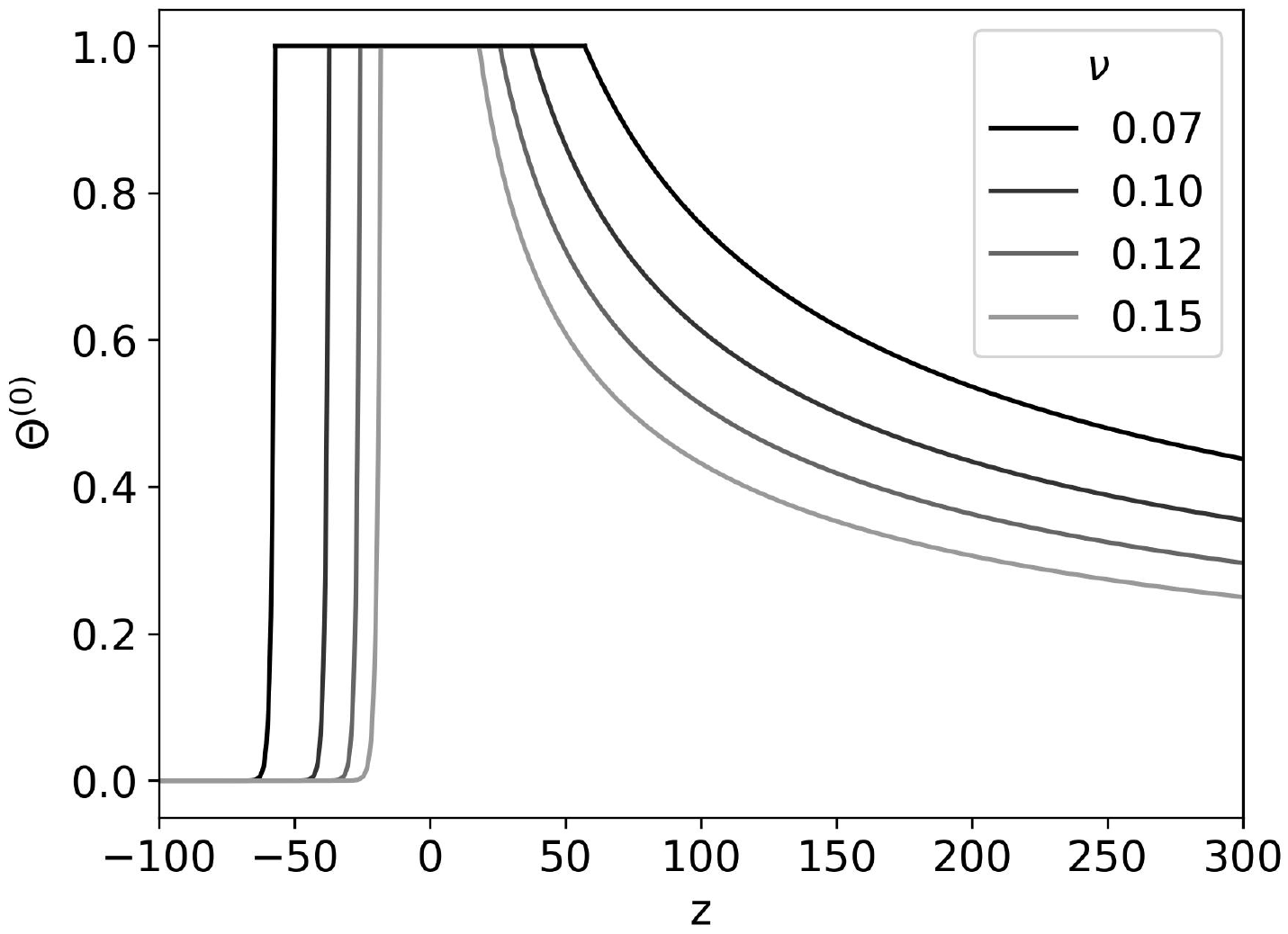}
	\caption{Profiles of $\Theta^{(0)}$ for $\alpha=1$ and different values of $\nu$ for global losses. Upper branch.}
	\label{fig:tehg1}
\end{figure}

\begin{figure}[p]
	\centering\captionsetup{justification=centering}
	\includegraphics[width=0.49\columnwidth]{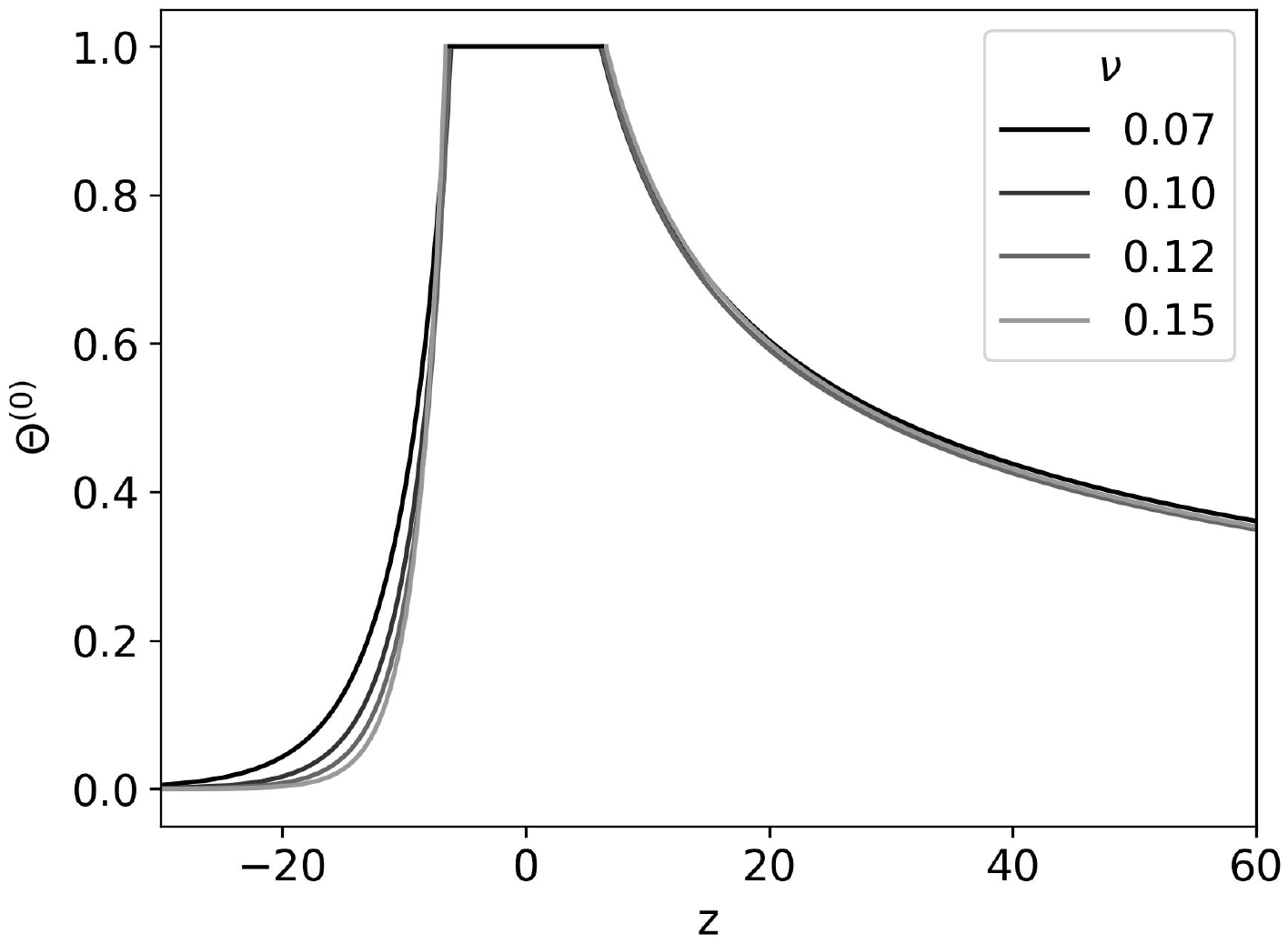}
	\caption{Profiles of $\Theta^{(0)}$ for $\alpha=1$ and different values of $\nu$ for global losses. Lower branch.}
	\label{fig:teh1gl}
\end{figure}

\begin{figure}[p]
	\centering\captionsetup{justification=centering}
	\includegraphics[width=0.49\columnwidth]{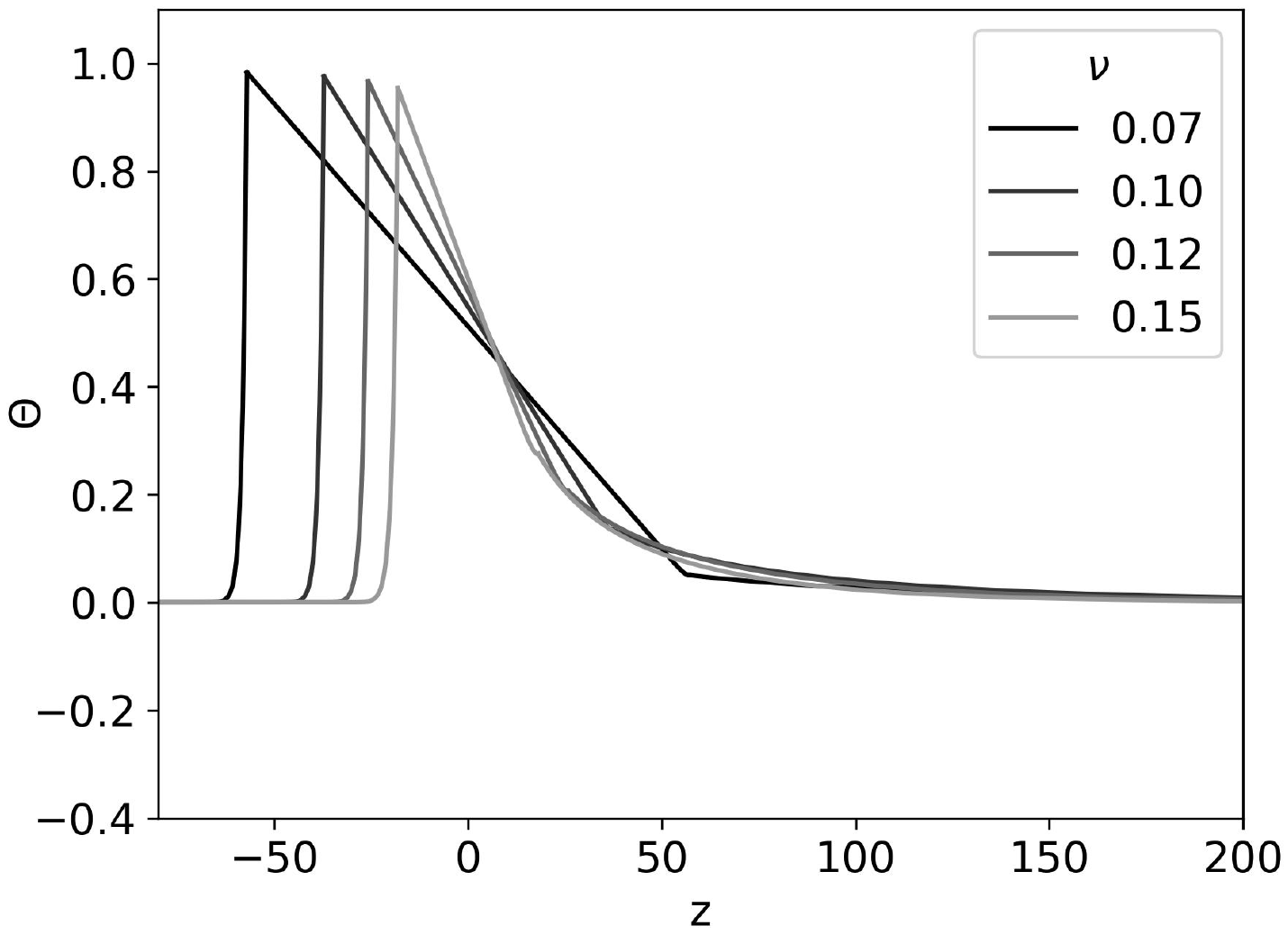}
	\caption{Profiles of $\Theta$ for $\alpha=1$, $\beta=10$ and different values of $\nu$ for global losses. Upper branch. Uniform approximation.}
	\label{fig:teh1ga}
\end{figure}

\begin{figure}[p]
	\centering\captionsetup{justification=centering}
	\includegraphics[width=0.49\columnwidth]{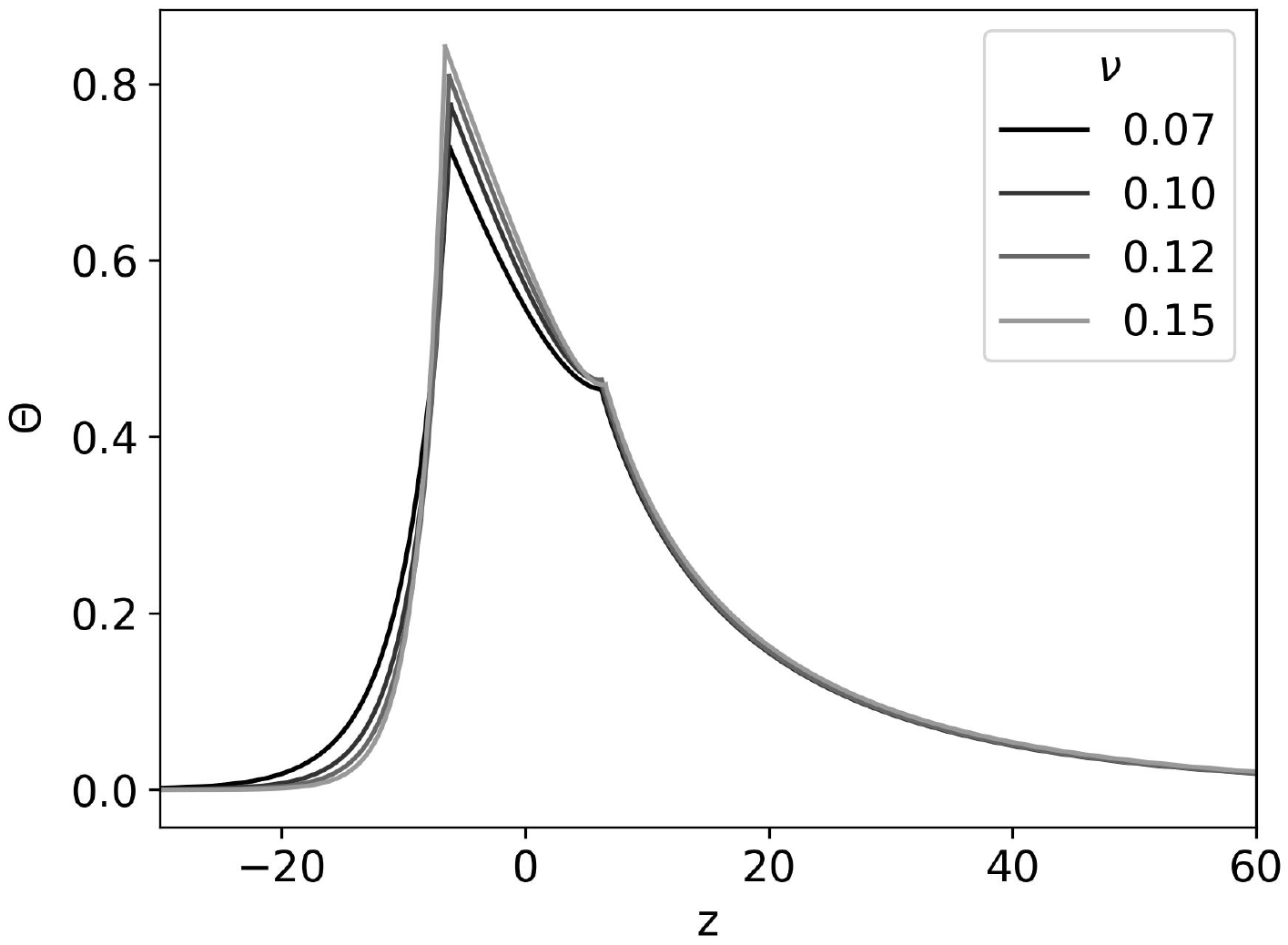}
	\caption{Profiles of $\Theta$ for $\alpha=1$, $\beta=10$ and different values of $\nu$  for global losses. Lower branch. Uniform approximation.}
	\label{fig:teh1lga}
\end{figure}

\begin{figure}[p]
	\centering\captionsetup{justification=centering}
	\includegraphics[width=0.49\columnwidth]{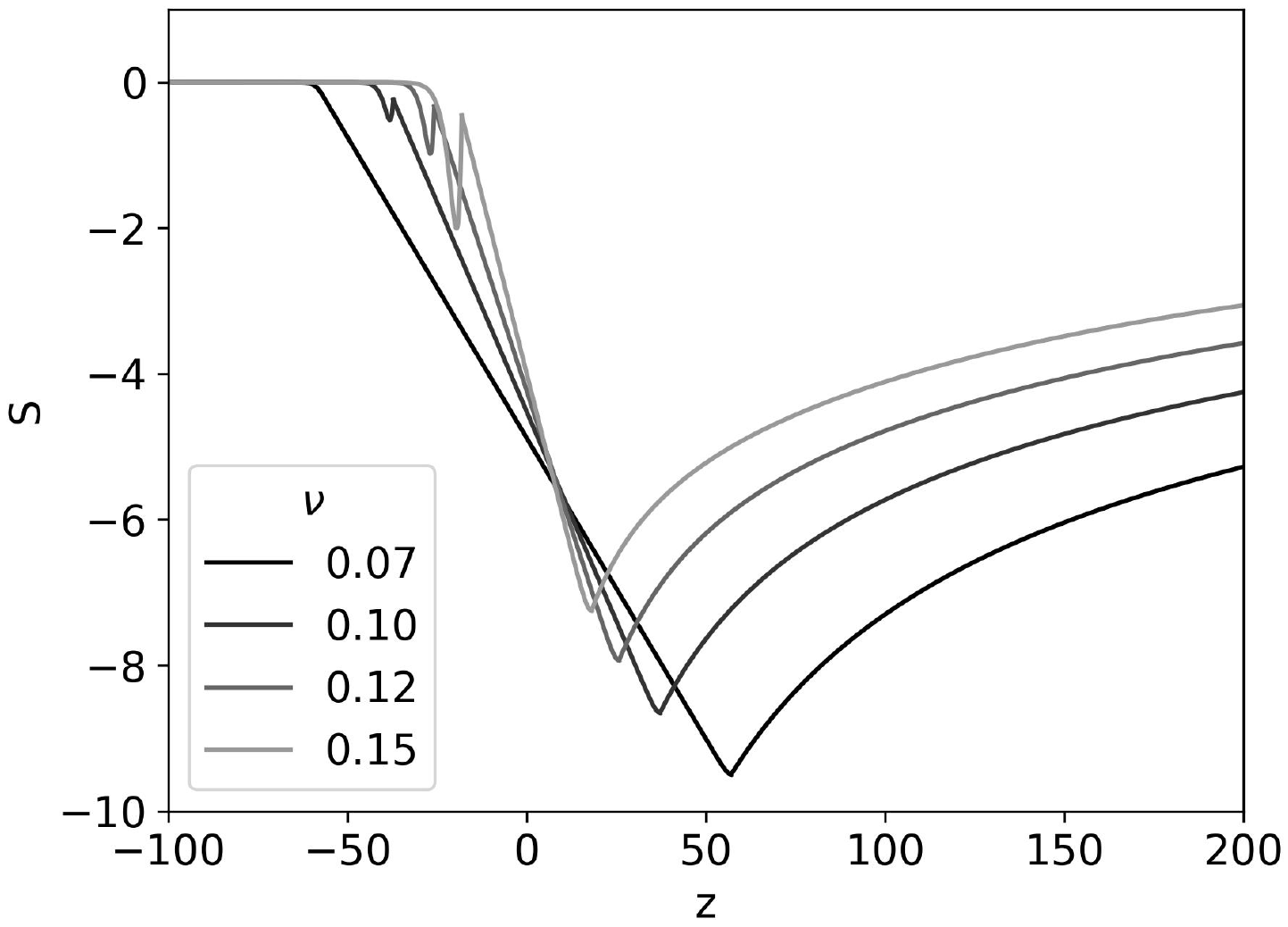}
	\caption{Profiles of $S$ for $\alpha=1$ and different values of $\nu$ for global losses. Upper branch. Uniform  approximation.}
	\label{fig:Sga}
\end{figure}

\begin{figure}[p]
	\centering\captionsetup{justification=centering}
	\includegraphics[width=0.49\columnwidth]{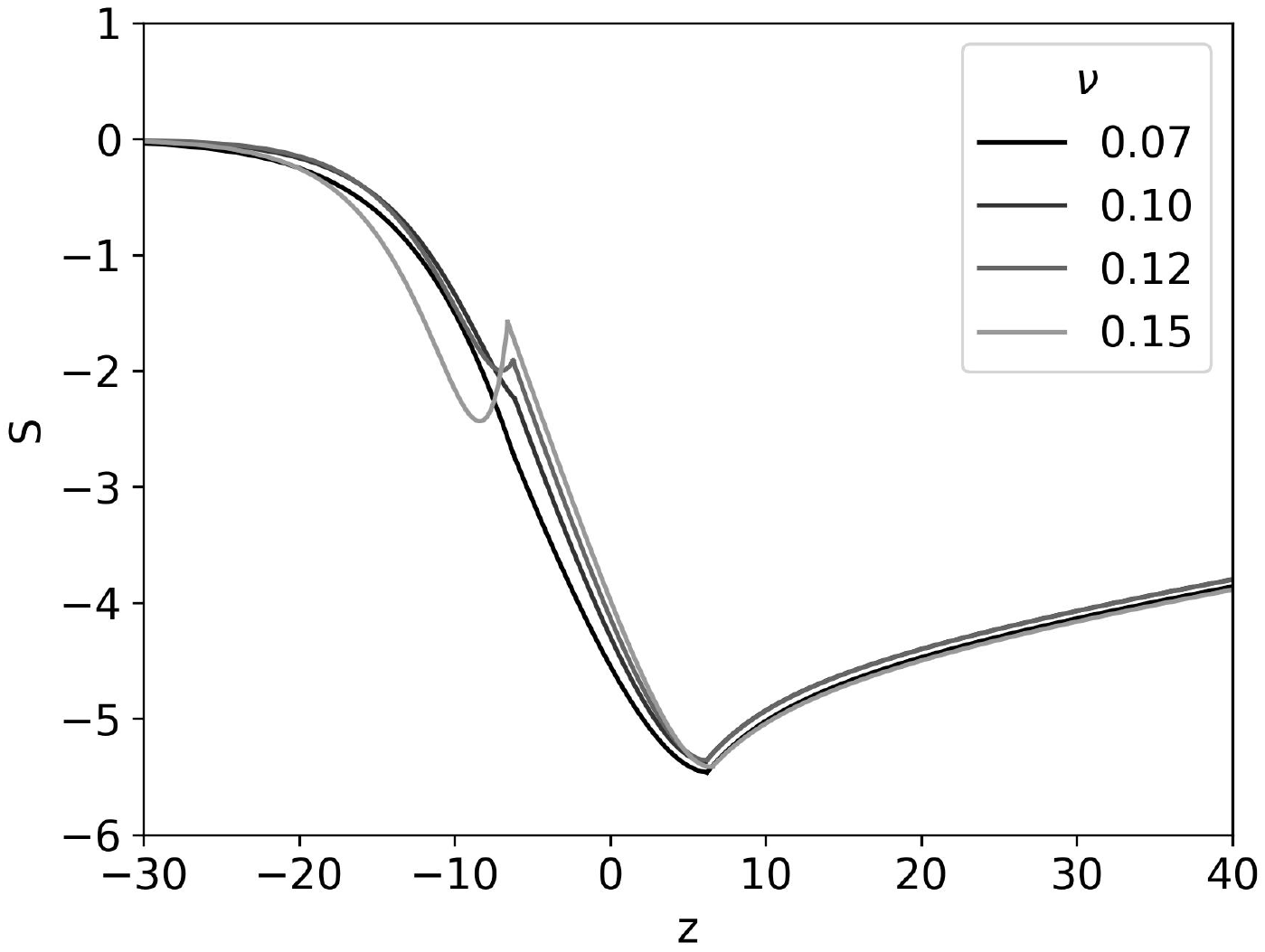}
	\caption{Profiles of $S$ for $\alpha=1$ and different values of $\nu$   for global losses. Lower branch. Uniform  approximation.}
	\label{fig:S1gl}
\end{figure}

The profiles for global heat losses are shown in Figures \ref{fig:tehg1} to \ref{fig:S1gl}.





\afterpage{\clearpage}

\newpage

\section{Conclusions}

This article constitutes a theoretical contribution to the studies of the flame front disintegration in Hele-Shaw cells \cite{veiga2020unexpected, escanciano2021characterization, martinez2019role, dominguez2022stable}. The adopted approach is conceptually linked to the theory of Zeldovich et al. \cite{zeldovich1985mathematical} of flame-balls stabilized
by heat losses. The phenomenon is explored employing a simplified 1D model. As a result we were able to calculate the radius-velocity-Lewis number-heat losses locus allowing for 2D self-drifting flame-balls.

Two branches of small-large radius, and low-high velocity are revealed for the same level of heat losses --aka gap between Hele-Shaw plates. 

The 2D flame-balls have a fundamental characteristic which distinguishes them from their 3D counterparts. Specifically, one of the principal outcomes of \cite{brailovsky1997stationary} is that a 3D self-drifting flame-ball emerges as a bifurcation from a steady spherical structure. This bifurcation was later confirmed by Minaev et al. \cite{minaev2001self, minaev2002self, kagan2004self} through a rigorous mathematical analysis of a 3D model.

The lack of a steady solution and a bifurcation is one of the basic differences between the 3D studies and the results of our analysis dealing with 2D flamelets.

\GS{The present work is devoted to estimates of the propagation velocity of an individual flame ball in a flat channel.  It would be interesting to extend the analysis over the collective propagation of a group of flame balls as appears in Figs. \ref{fig:abb_gen2}-\ref{fig:abb_gen4}, and where the flamelets compete for the common fuel and mutual heating of one another.  A mean-field type of approach as that developed by D'Angelo and Joulin \cite{d2001collective} or Willimas and Grcar \cite{williams2009hypothetical} seems particularly promising.}

In our understanding of the phenomenon, flame-balls emerge in gaseous premixtures as an extreme case of diffusive-thermal instability where invariably $Le<1$. A similar pattern is observed in smoldering burning of thin solid sheets with and without flows \cite{zhang1992lewis, zik1998fingering, zik1999fingering, matsuoka2017appearance, zhu2019microgravity, malchi2007nano, olson1998finger, uchida2015experimental}
. It would be interesting to extend the 1D approach of this paper also to the above gas-solid configurations.

\section*{Acknowledgements}
This research was supported in part by the US-Israel Binational Science Foundation (Grant 2020-005).


\end{document}